\documentclass[12pt]{article}

\usepackage{graphicx}
\usepackage{dcolumn}
\usepackage{bm}
\usepackage[a4paper,top=3.5cm,bottom=3.2cm,left=2.3cm,right=2.3cm,bindingoffset=5mm]{geometry}
\usepackage{amsfonts}
\usepackage{latexsym}
\usepackage{amsmath}
\usepackage{amssymb}
\usepackage[english]{babel}
\usepackage[latin1]{inputenc}
\usepackage[T1]{fontenc}
\usepackage{slashed}
\usepackage{float}
\usepackage{subfigure}
\usepackage{color}

    \newcommand{\be}{\begin{equation}}
    \newcommand{\ee}{\end{equation}}
    \newcommand{\ba}{\begin{eqnarray}}
    \newcommand{\ea}{\end{eqnarray}}
    \newcommand{\eite}{\end{itemize}}
    \newcommand{\bite}{\begin{itemize}}

\def\gtrsim{\ \raisebox{-.4ex}{\rlap{$\sim$}} \raisebox{.4ex}{$>$}\ }

\begin{document}

\hfill{{\small Ref. SISSA 23/2018/FISI}}

\hfill{{\small Ref. IPMU18--0112}

\hfill{{\small Ref. IPPP/18/47}}

\vspace{1.0cm}
\begin{center}
{\bf{\large On Neutrino Mixing in Matter and CP and T Violation 
Effects in Neutrino Oscillations
}}\\

\vspace{0.4cm}
S. T. Petcov$\mbox{}^{a,b)}$
\footnote{Also at: Institute of Nuclear Research and
Nuclear Energy, Bulgarian Academy of Sciences, 1784 Sofia, Bulgaria}
and Ye-Ling Zhou$\mbox{}^{c)}$ 

\vspace{0.1cm}
$\mbox{}^{a)}${\em  SISSA/INFN, Via Bonomea 265, 34136 Trieste, Italy.\\}

\vspace{0.1cm}
$\mbox{}^{b)}${\em Kavli IPMU (WPI), The University of Tokyo, Kashiwa,
Chiba 277-8583, Japan.\\}

\vspace{0.1cm}
$\mbox{}^{c)}${\em  Institute for Particle Physics Phenomenology, 
Department of Physics, Durham University, Durham DH1 3LE,
United Kingdom.}

\end{center}

\begin{abstract}
Aspects of 3-neutrino mixing and oscillations in vacuum 
and in matter with constant density are investigated 
working with a real form of the neutrino Hamiltonian. 
We find the (approximate) equalities 
$\theta^m_{23} = \theta_{23}$ and $\delta^m = \delta$,
$\theta_{23}$ ($\theta^m_{23}$) and $\delta$ ($\delta^m$)
being respectively the atmospheric neutrino 
mixing angle and the Dirac CP violation phase  
in vacuum (in matter) 
of the neutrino mixing matrix, which 
are shown to represent excellent approximations 
for the conditions of the 
T2K (T2HK), T2HKK, NO$\nu$A and DUNE  
neutrino oscillation experiments.
A new derivation of the known relation 
$\sin2\theta^m_{23} \sin\delta^m 
= \sin2\theta_{23} \sin\delta$ is presented 
and it is used to obtain a correlation between 
the shifts of $\theta_{23}$ and $\delta$
due to the matter effect.
A derivation  of the relation between 
the rephasing invariants
which determine the magnitude of CP and T violating effects 
in 3-flavour neutrino  oscillations 
in vacuum, $J_{\rm CP}$, and of the T violating effects
in matter with constant density, $J^{m}_{\rm T} \equiv J^{m}$, 
reported in \cite{Krastev:1988yu} without a proof,
is presented. 
It is shown that the function $F$ 
which appears in this relation, $J^{m} = J_{\rm CP}\,F$,
and whose explicit form 
was given in \cite{Krastev:1988yu}, 
coincides with the function $\tilde{F}$ in the similar relation 
$J^{m} = J_{\rm CP}\tilde{F}$ derived in \cite{Naumov:1991ju},
although  $F$ and $\tilde{F}$ 
are expressed in terms of different sets of 
neutrino mass and mixing parameters and 
have completely different forms.

\end{abstract}

\baselineskip 12pt

\section{Introduction and Preliminary Remarks} 
\vspace{-0.2cm}
It was shown in 1988 in ref. \cite{Krastev:1988yu} that
in the case of what is currently referred to as the reference 3-neutrino mixing
(see, e.g., \cite{PDG2017}), the magnitude of the CP and T 
 violating (T violating) effects 
 in  neutrino oscillations in vacuum 
(in matter with constant density) are 
controlled by the rephasing invariant  $J_{\rm CP}$  
($J^m_{\rm T} \equiv J^m$) 
associated with the Dirac CP violation phase present in the  
Pontecorvo, Maki, Nakagawa and Sakata (PMNS) \cite{BPont5767,MNS62}
neutrino mixing matrix:
%%%%%%%%%%%%%%%%%%%%%%%%%
\be
J_{\rm CP} (J^m)  =  {\rm Im}\left( \left(U^{(m)}_{e2}\right) 
\left(U^{(m)}_{\mu 3}\right) \left(U^{(m)}_{e3}\right)^\ast 
\left(U^{(m)}_{\mu 2}\right)^\ast \right)\,,
\label{JCPvm}
\ee
%%%%%%%%%%%%%%%%%%%%%%%%%%%%%%
%
where $U^{(m)}_{li}$, $l=e,\mu,\tau$, $i=1,2,3$, 
are the elements of the PMNS matrix in vacuum (in matter) 
$U^{(m)}$. The CP violating asymmtries in the case of neutrino 
oscillations in vacuum, for example,
%%%%%%%%%%%%%%%%%%%%%%%%%%%%%%%%%%%%%%%%
\begin{eqnarray}
A^{(l,l')}_{\rm CPvac} = P^{\rm vac}({\nu_l \rightarrow \nu_{l'}}) 
- P^{\rm vac}({\bar{\nu}_l \rightarrow \bar{\nu}_{l'}})\,,~~
l\neq l'\quad\text{and}\quad l,\,l'=e,\mu,\tau\,,
\end{eqnarray}
%%%%%%%%%%%%%%%%%%%%%%%%%%%%%%%%%%%%%%%%%%
%
$P^{\rm vac}({\nu_l \rightarrow \nu_{l'}})$ 
and $ P^{\rm vac}({\bar{\nu}_l \rightarrow \bar{\nu}_{l'}})$ 
being the probabilities of respectively
$\nu_l \rightarrow \nu_{l'}$ 
and $\bar{\nu}_l \rightarrow \bar{\nu}_{l'}$
oscillations, were shown to be given 
by \cite{Krastev:1988yu}:
%%%%%%%%%%%%%%%%%%%%%%%%%%%%%%%%%%%%
\begin{equation}
\label{CPTAsym}
A^{(e,\mu)}_{\rm CPvac} = A^{(\mu,\tau)}_{\rm CPvac} = 
- A^{(e,\tau)}_{\rm CPvac} = 4\,J_{\rm CP}~\Phi^{\rm vac}_{\rm osc}\,,
\end{equation}
%%%%%%%%%%%%%%%%%%%%%%%%%%%%
%
with
%%%%%%%%%%%%%%%%%%%%%%%%%%%%%%%%%%%%%%%%%
\begin{equation}
\label{JCPF}
\Phi^{\rm vac}_{\rm osc} =
\sin\left(\frac{\Delta m^2_{21}L}{2E}\right) +
\sin\left(\frac{\Delta m^2_{32}L}{2E}\right) +
\sin\left(\frac{\Delta m^2_{13}L}{2E}\right)\,.
\end{equation}
%%%%%%%%%%%%%%%%%%%%%%%%%%%%%%%%%%
%
where $\Delta m^2_{ij} = m^2_i - m^2_j$, $i\neq j$, $m_i$, 
$i=1,2,3$, is the mass of the neutrino $\nu_i$ 
with definite mass in vacuum,
$E$ is the neutrino energy and $L$ is the distance travelled by 
the neutrinos. In \cite{Krastev:1988yu} similar results were shown 
to be valid for the T-violating asymmetries  
in oscillations in vacuum (in matter),
$A^{(l',l)}_{\rm Tvac(m)} = P^{vac(m)}({\nu_l \rightarrow \nu_{l'}}) 
- P^{vac(m)}({\nu_{l'} \rightarrow \nu_{l}})$:\\
%%%%%%%%%%%%%%%%%%%%%%%%%%%%%%%%%%%%%%%%%%%
\begin{equation}
A^{(\mu,e)}_{\rm Tvac(m) } = A^{(\tau,\mu)}_{\rm Tvac(m)} 
= -\,A^{(e,\tau)}_{\rm Tvac(m)} = 
4\,J^{(m)}_{\rm CP(T)}~\Phi^{\rm vac(m)}_{\rm osc}\,,
\label{eq:AT}
\end{equation}
%%%%%%%%%%%%%%%%%%%%%%%%%%%%%%%%%%%%%
%
where $\Phi^{\rm m}_{\rm osc}$ has the same form as 
$\Phi^{\rm vac}_{\rm osc}$ in eq. (\ref{JCPF}) with 
$\Delta m^2_{ij}$ replaced by mass splitting in the matter $\Delta M^2_{ij}=M^2_i - M^2_j$,
and $M_i$, $i=1,2,3$, are neutrino mass-eigenvalues in the matter. In vacuum 
the T violating asymmetries in antineutrino oscillations,
$\bar{A}^{(l',l)}_{\rm Tvac} = 
P^{\rm vac}({\bar{\nu}_l \rightarrow \bar{\nu}_{l'}}) 
- P^{\rm vac}({\bar{\nu}_{l'} \rightarrow \bar{\nu}_{l}})$, 
are related to  those in neutrino oscillations
owing to the CPT invariance:
$\bar{A}^{(l',l)}_{\rm Tvac} = -\,A^{(l',l)}_{\rm Tvac}$.
In ordinary matter (Earth, Sun) 
\footnote{By ``ordinary'' we mean matter which does 
not contain antiprotons, antineutrons and positrons.}
the presence of matter causes CP and CPT  violating 
effects in neutrino oscillations \cite{Lang87} 
and $|\bar{A}^{(l',l)}_{\rm Tm} |\neq |A^{(l',l)}_{\rm Tm}|$.
However, in ordinary  matter with constant density or 
with density profile which is symmetric relative to the middle point, 
like the matter of the Earth, the matter effects
preserve the T symmetry and do not generate 
T violating effects in neutrino oscillations
\cite{Krastev:1988yu}. Thus, T violating effects in 
the flavour neutrino oscilations 
taking place when the neutrinos traverse, e.g., the Earth mantle 
or the Earth core can be caused in the case of 3-neutrino mixing  
only by the Dirac phase in the PMNS matrix.

 The  $J_{\rm CP}$-factor in the expressions 
for  $A^{(l,l')}_{\rm CP(T)~vac}$, $l\neq l'$,   
is analogous to the rephasing invariant  
associated with the Dirac phase in the Cabibbo-Kobayashi-Maskawa (CKM)
quark mixing matrix, introduced in \cite{Jarlskog:1985ht}.
In the standard parametrization of the
PMNS mixing matrix (see, e.g., \cite{PDG2017})
it has the form:
%%%%%%%%%%%%%%%%%%%%%%%%%%%%%%
 \begin{equation}
 \begin{split}
 J_{\rm CP}= & \; 
 \frac{1}{8}\,\cos\theta_{13}
 \sin 2\theta_{12}\,\sin 2\theta_{23}\,\sin 2\theta_{13}\,\sin \delta\,,
 \label{JCPstpar}
 \end{split}
 \end{equation}
%%%%%%%%%%%%%%%%%%%%%%%%%%%%%%%
% 
where $\theta_{12}$, $\theta_{23}$ and $\theta_{13}$ are the solar, 
atmospheric and reactor neutrino mixing angles 
and $\delta$ is the Dirac CP violation phase.
The expression for the $J_{\rm CP}$-factor
is the same in the parametrisation
of the PMNS matrix $U_{\rm PMNS}\equiv U$
employed in  \cite{Krastev:1988yu}:
%%%%%%%%%%%%%%%%%%%%%%%%%%%%%%%%%
\begin{equation}
  U = R_{23}(\theta_{23})\,P_{33}(\delta)\,R_{13}(\theta_{13})\,
  R_{12}(\theta_{12})\,,
\label{eq:PMNSKP1}
\end{equation}
%%%%%%%%%%%%%%%%%%%%%%%%%%%%%%%%%%
%
where
%%%%%%%%%%%%%%%%%%%%%%%%%%%%%%%%%%%%%%%%%%%
\begin{equation}
  R_{23}\left(\theta_{23} \right) =
  \begin{pmatrix}
1 & 0 & 0\\
0 & \cos \theta_{23} & \sin \theta_{23} \\
0 & - \sin \theta_{23}  & \cos \theta_{23} 
\end{pmatrix} \,,
\quad
 P_{33}(\delta) = {\rm diag}(1,1,e^{i\delta})\,,
\label{eq:R23P33}
\end{equation}
%%%%%%%%%%%%%%%%%%%%%%%%%%%
%
and
%%%%%%%%%%%%%%%%%%%%%%%%%%%
\begin{equation}
R_{13}\left(\theta_{13} \right) = \begin{pmatrix}
\cos \theta_{13} & 0 & \sin \theta_{13}\\
0 & 1 & 0\\ 
- \sin \theta_{13} & 0 & \cos \theta_{13}
\end{pmatrix} \,,
\quad
R_{12}\left(\theta_{12} \right) = \begin{pmatrix}
\cos \theta_{12} & \sin \theta_{12} & 0\\
- \sin \theta_{12} & \cos \theta_{12} & 0\\
0 & 0 & 1 \end{pmatrix} \,.
\label{R122313}
\end{equation}
%%%%%%%%%%%%%%%%%%%%%%%%%
%
The expression of the PMNS matrix in the 
standard parametrisation 
$U^{sp}$, is related to the expression in the 
parametrisation in eq. (\ref{eq:PMNSKP1})
as follows: $U^{sp} = UP^*_{33}(\delta)$.

 In eq. (\ref{eq:PMNSKP1}) the two CP violation (CPV) 
Majorana phases present in $U_{\rm PMNS}$ in the case of 
massive Majorana neutrinos \cite{Bilenky:1980cx}
were omitted since, as was shown in \cite{Bilenky:1980cx,Lang87}, 
the probabilities of flavour neutrino oscillations 
of interest for the study performed in \cite{Krastev:1988yu} 
and for the present study, do not depend on the Majorana phases. 
Thus, the results presented in \cite{Krastev:1988yu} and 
the new results derived in the present article are valid 
for both Dirac and Majorana neutrinos 
with definite masses in vacuum.

 In ref. \cite{Krastev:1988yu} the following relation between 
the rephasing invariants in vacuum and in matter with constant 
density, $J_{\rm CP}$ and  $J^m_{\rm T}\equiv J^m$, has been reported:
%%%%%%%%%%%%%%%%%%%
\be
J^m = J_{\rm CP}\,
F(\theta_{12},\theta_{13},\Delta m^2_{21},\Delta m^2_{31},A)\,,
\label{eq:JCPmJCPv}
\ee
%%%%%%%%%%%%%%%%%%%%%%%%
%
where $A=2E\,\sqrt{2}\, G_{\rm F}\,N_e$ is the matter term 
\cite{Matter1,Matter2,Matter3},
$G_{\rm F}$ and $N_e$ being respectively 
the Fermi constant and the electron number density of matter. 
The function $F$
in eq. (\ref{eq:JCPmJCPv})
was given in the following explicit form in  \cite{Krastev:1988yu}:
%%%%%%%%%%%%%%%%%%%%%%%%%%%%%%%%%%%%
\begin{equation}
F = \frac{F_1}{F_2\,F_3}\,D_{12}\, D_{13}\,D_{23}\,D_{32}\,, 
\label{eq:F}
\end{equation}
%%%%%%%%%%%%%%%%%%%%%%%%%%%%%%
%
where
%%%%%%%%%%%%%%%%%%
\begin{equation}
D_{ij} \equiv  m^2_i - M^2_j\,,~~i,j=1,2,3\,,  
\label{Dij}
\end{equation}  
%%%%%%%%%%%%%%%%%%%%%  
%
%%%%%%%%%%%%%%%%%%%%%%%
\begin{eqnarray}
  F_1 = D_{12}D_{13}D_{23}D_{32} &+
  A\,\left [ D_{13}D_{23}\left (D_{32} - \Delta m^2_{31}\,|U_{e3}|^2\right )
+ D_{12}D_{32}\left (D_{23} - \Delta m^2_{21}\,|U_{e2}|^2\right )\right ]
\nonumber
\\[0.3cm]
&+ A^2\left [|U_{e1}|^2\,D_{32}\,D_{23} + |U_{e2}|^2\,D_{32}\,D_{13} +
    |U_{e3}|^2\,D_{12}\,D_{23}\right ]\,,
\label{eq:F1}
\end{eqnarray}
%%%%%%%%%%%%%%%%%%%%%%%%%%%%%%%
%
%%%%%%%%%%%%%%%%%%%%%%%
\begin{equation}
  F_2 = |U_{e1}|^2\left (D_{12} + A \right )^2D^2_{32}
+ |U_{e2}|^2D^2_{12}D^2_{32}
+ |U_{e3}|^2D^2_{12}\left (D_{32} + A \right )^2 
- A^2\,|U_{e1}|^2|U_{e3}|^2(\Delta m^2_{31})^2\,,
\label{eq:F2}
\end{equation}
%%%%%%%%%%%%%%%%%%%%%%%%%%%%%%%
%
%%%%%%%%%%%%%%%%%%%%%%%%%%
\begin{equation}
F_3 = |U_{e1}|^2\left(D_{13} + A \right )^2D^2_{23}
+ |U_{e3}|^2 D^2_{13}D^2_{23}
+ |U_{e2}|^2D^2_{13}\left (D_{23} + A\right )^2 
- A^2\,|U_{e1}|^2\,|U_{e2}|^2(\Delta m^2_{21})^2\,.
\label{eq:F3}
\end{equation}
%%%%%%%%%%%%%%%%%%%%%%%%%%%%%%
%

As was noticed in  \cite{Krastev:1988yu},
the function $F_3$ can formally be obtained from the function $F_2$
by interchanging $m^2_2$ and $m^2_3$, $M^2_2$ and $M^2_3$, and
$|U_{e2}|^2$ and $|U_{e3}|^2$. 
In the parametrisation (\ref{eq:PMNSKP1}) 
used in \cite{Krastev:1988yu} and, thus in eqs. (\ref{eq:F1}) - (\ref{eq:F3}),
$U_{e1}$, $U_{e2}$ and $U_{e3}$ are real quantities:
$U_{e1} = c_{12}c_{13}$, $U_{e2} =s_{12}c_{13}$ and $U_{e3}=s_{13}$,
where $c_{ij} \equiv \cos\theta_{ij}$ and  $s_{ij} \equiv \sin\theta_{ij}$.
Thus, $|U_{ei}|^2 =U^2_{ei}$, $i=1,2,3$. 
The function $F(\theta_{12},\theta_{13},\Delta m^2_{21},\Delta m^2_{31},A)$ 
as defined by eqs. (\ref{eq:F}) - (\ref{eq:F3}), 
depends, in particular, on the differences between 
the squares of the neutrino masses in vacuum and in matter,
$D_{ij} = m^2_i - M^2_j$, $i\neq j$.
However, as it follows from the form of the Hamiltonian 
of the neutrino system in matter with constant density, 
whose eigenvalues are  $M^2_j/(2E)$ (see further), as well as from 
the explicit analytic expressions for $M^2_j$ derived 
in \cite{Matter2},
the mass squared differences $D_{ij}$ of interest are functions 
of  $\theta_{12}$, $\theta_{13}$, $\Delta m^2_{21}$, $\Delta m^2_{31}$ 
and $A$ and do not depend on $\theta_{23}$ and $\delta$.
As a consequence, the function $F$ in eq. (\ref{eq:JCPmJCPv})
is independent on $\theta_{23}$ and $\delta$ \cite{Krastev:1988yu}:
$F = F(\theta_{12},\theta_{13},\Delta m^2_{21},\Delta m^2_{31},A)$.
 
 In deriving the relation (\ref{eq:JCPmJCPv}), the following 
parametrisation of the neutrino mixing matrix in matter $U^m$  
was used:
%%%%%%%%%%%%%%%%%%%%%%%%%%%%%%%%%
\begin{equation}
  U^m = Q\,R_{23}(\theta^m_{23})\,P_{33}(\delta^m)\,R_{13}(\theta^m_{13})\,
  R_{12}(\theta^m_{12})\,,~Q={\rm diag}(1,e^{i\beta_2},e^{i\beta_3})\,,
\label{eq:PMNSKP1m}
\end{equation}
%%%%%%%%%%%%%%%%%%%%%%%%%%%%%%%%%%
%
where $\theta^m_{23}$, $\theta^m_{13}$, $\theta^m_{12}$, 
$\delta^m$ are the neutrino mixing angles and the Dirac 
CPV phase in matter and the Majorana CPV phases 
were omitted. The phases $\beta_2$ and $\beta_3$ 
in the matrix $Q$ are unphysical and do not play any role 
in the derivation of relation (\ref{eq:JCPmJCPv}).
They ensure that the matrix $U^m$ can be cast in the 
form given in eq. (\ref{eq:PMNSKP1m}) \cite{Zaglauer:1988gz}
(see also \cite{Freund:2001pn}).
Obviously, the parametrisation of $U^m$ in  eq. (\ref{eq:PMNSKP1m}) 
is analogous to the parametrisation (\ref{eq:PMNSKP1}) 
of the neutrino mixing matrix in vacuum.

It follows from eqs. (\ref{eq:F}) - (\ref{eq:F3}) that 
 \cite{Krastev:1988yu} in the case of oscillations 
in vacuum, i.e., for $N_e = 0$ ($A=0$), one has 
%%%%%%%%%%%%%%%%%%%%%%%%
\begin{equation}
F(\theta_{12},\theta_{13},\Delta m^2_{21},\Delta m^2_{31},0) = 1\,,
\label{eq:FA0}  
\end{equation}
%%%%%%%%%%%%%%%%%%%%%%%
%
and that $F$ is symmetric with respect to the interchange of
$m^2_2$ and $m^2_3$, $M^2_2$ and $M^2_3$ and of
$|U_{e2}|^2$ and $|U_{e3}|^2$. 

 The relation (\ref{eq:JCPmJCPv}) between 
$J^m$ and $J_{\rm CP}$ implies, in particular, 
that we can have $J^m\neq 0$ only if 
$J_{\rm CP}\neq 0$, i.e., 
T violation effects can be present 
in neutrino oscillations taking place in matter 
with constant density or density distributed symmetrically relative to the middle point (like in the Earth)
only if CP and T violation effects are present 
in neutrino oscillations taking place in vacuum. 
It was shown also in \cite{Krastev:1988yu}
that  the presence of matter can enhance somewhat   
$|J^m|$ with respect to 
its vacuum value $|J_{\rm CP}|$: 
in the example considered in \cite{Krastev:1988yu} 
the enhancement was by a factor of 3.
Taking the best fit values of neutrino oscillation parameters 
for neutrino mass spectrum with normal ordering (inverted ordering)
\footnote{For a discussion of the different possible types 
of neutrino mass spectrum see, e.g., \cite{PDG2017}.}
obtained in the global analysis in \cite{Esteban:2016qun},  
%%%%%%%%%%%%%%%%%%%%%%%%%%%%
\begin{eqnarray}
\theta_{12} =33.62^\circ\,, ~ \theta_{23} = 47.2^\circ\, (48.1^\circ)\,,
~ \sin^2\theta_{13} = 8.54^\circ\, (8.58^\circ)\,,
~\delta = 234^\circ\, (278^\circ)\,, 
\nonumber\\
\Delta m^2_{21} = 7.4 \times 10^{-5} {\rm eV^2} \,,
~ \Delta m^2_{31} = 2.494 \times 10^{-3} {\rm eV^2}\, 
(\Delta m^2_{32} = -2.465 \times 10^{-3} {\rm eV^2})\,,
\label{eq:bfvnoscparam}
\end{eqnarray}
%%%%%%%%%%%%%%%%%%%%%%%%%%%%%%
% 
one always has for the ratio $|J^m/J_{\rm CP}| < 1.2$ \cite{Xing:2018lob}.
This result persists even if we fix $\delta$ to its best fit value 
and vary the other neutrino oscillation parameters in 
their $3\sigma$ allowed ranges determined in \cite{Esteban:2016qun}.
Relaxing arbitrarily the $3\sigma$ experimental constraints on 
the allowed ranges of 
$\Delta m^2_{21}$ and $\Delta m^2_{31}$, we find  
that indeed the maximal enhancement factor $|J^m/J_{\rm CP}|$ 
is 3.6 for neutrino mass spectrum with normal ordering (NO) and 
2.9 for spectrum with inverted ordering (IO). In both cases, 
the maximal enhancement corresponds to $J^m$ reaching 
its theoretical maximal value ${\rm max}(|J^m|) = 1/(6\sqrt{3})$.

 In 1991 in \cite{Naumov:1991ju} a relation similar to that given in 
eq. (\ref{eq:JCPmJCPv}) was obtained:
%%%%%%%%%%%%%%%%%%%%%%%%%%
\begin{equation}
J^m = J_{\rm CP}\,\tilde{F}\,.
\label{eq:JCPmJCPv2}
\end{equation}
%%%%%%%%%%%%%%%%%%%%%%%%
%
The function $\tilde{F}$ was given in the following form:
%%%%%%%%%%%%%%%%%%%%%%%%%%%%%
\begin{equation}
\tilde{F} = \frac{\Delta m^2_{12}\,\Delta m^2_{23}\,\Delta m^2_{31}}
{\Delta M^2_{12}\,\Delta M^2_{23}\,\Delta M^2_{31}}\,,
\label{eq:tildeF}
\end{equation} 
%%%%%%%%%%%%%%%%%%%%%%%%%%%%%%%
%              
where 
% $\Delta m^2_{ij} = m^2_i - m^2_j$ and 
$\Delta M^2_{ij} =  M^2_i - M^2_j$.

 The relation (\ref{eq:JCPmJCPv}) 
between the rephasing invariants 
$J^m$ and $J_{\rm CP}$
was presented in \cite{Krastev:1988yu} without a proof. 
In the present article, after discussing certain aspects of 
neutrino mixing in matter,  
we provide a derivation of 
the relation (\ref{eq:JCPmJCPv}). Further, we show that 
the function $F$ in eq. (\ref{eq:JCPmJCPv}), as defined in 
eqs. (\ref{eq:F1}) -  (\ref{eq:F3}), coincides 
with the function $\tilde{F}$ 
in the relation (\ref{eq:JCPmJCPv2}) obtained in 
\cite{Naumov:1991ju},
%%%%%%%%%%%%%%%%%%%%%%%%%%%%%
\begin{equation}
F = \tilde{F}\,,
\label{eq:FeqFt}
\end{equation}
%%%%%%%%%%%%%%%%%%%%%%%%%%%%%
% 
i.e., that the function $F$ 
is just another representation of the function $\tilde{F}$.

%%%%%%%%%%%%%%%%%%%%%%%%%%%%%%%%%%
%
\section{On the 3-Neutrino Mixing in Matter} 
%
%%%%%%%%%%%%%%%%%%%%%%%%%%%%%
 
In  \cite{Krastev:1988yu} the analysis was performed starting with the 
following Hamiltonian of the neutrino system in matter diagonalised 
with the help of the neutrino mixing matrix in matter $U^m$ \cite{Lang87}:
%%%%%%%%%%%%%%%%%%%%%%%%%%%%%
\begin{equation}
\frac{1}{2E} U\,
\left [ 
\begin{pmatrix}
m^2_1 & 0 & 0\\
0 & m^2_2 & 0\\
0 & 0 & m^2_3 
\end{pmatrix}
+
U^\dagger\,
\begin{pmatrix}
A & 0 & 0\\
0 & 0 & 0\\
0 & 0 & 0 
\end{pmatrix}\,U\right ]U^\dagger = 
\frac{1}{2E} U^m\,
\begin{pmatrix}
M^2_1 & 0 & 0\\
0 & M^2_2 & 0\\
0 & 0 & M^2_3 
\end{pmatrix}\, 
(U^m)^\dagger\,.
\label{eq:Hm} 
\end{equation}
%%%%%%%%%%%%%%%%%%%%%%%%%%%
%
It follows from the preceding equation 
\footnote{The CPV Majorana phases 
% \cite{Bilenky:1980cx}, 
$\alpha_{21}$ and $\alpha_{31}$, 
enter into the expression for the PMNS matrix 
in vacuum through the diagonal matrix 
$P = {\rm diag}(1,e^{i\alpha_{21}/2},e^{i\alpha_{31}/2})$ 
\cite{Bilenky:1980cx,BiPet87}):
$U_{\rm PMNS} = UP$. It follows from the expression 
in the left hand side of eq. (\ref{eq:Hm})
that the Hamiltonian of neutrino system in matter, 
and thus the 3-flavour neutrino oscillations in matter,
% in the case of interest  
do not depend on the Majorana phases \cite{Lang87}.
}
that the Hamiltonian of the neutrino system,
%%%%%%%%%%%%%%%%%%%%%%%%%%%%%
\begin{align}
H & = 
\dfrac{1}{2E}\,
% \Big[ 
\begin{pmatrix}
m^2_1 & 0 & 0\\
0 & m^2_2 & 0\\
0 & 0 & m^2_3 
\end{pmatrix}
+
\dfrac{1}{2E}\,  
U^\dagger\,
\begin{pmatrix}
A & 0 & 0\\
0 & 0 & 0\\
0 & 0 & 0 
\end{pmatrix}\,U\\
% \Big]
& =\dfrac{1}{2E}\,  
\begin{pmatrix}
m^2_1 + A|U_{e1}|^2 & AU^\ast_{e1}U_{e2}  & AU^\ast_{e1}U_{e3}  \\
AU^\ast_{e2}U_{e1}  & m^2_2 + A|U_{e2}|^2   &AU^\ast_{e2}U_{e3}  \\
AU^\ast_{e3}U_{e1} &AU^\ast_{e3}U_{e2}   & m^2_3 + A|U_{e3}|^2  
\end{pmatrix}
\label{eq:Hm3}
\end{align}
%%%%%%%%%%%%%%%%%%%%%%%%%%%%%%%%%%%%%%
%
is diagonalised by the matrix $U^\dagger U^m$ 
and its eigenvalues are $M^2_i/(2E)$, $i=1,2,3$.
In the parametrisation (\ref{eq:PMNSKP1}) of the PMNS matrix 
$U_{e1}$, $U_{e2}$ and $U_{e3}$ are real quantities:
$U_{e1} = c_{12}c_{13}$, $U_{e2} =s_{12}c_{13}$ and $U_{e3}=s_{13}$,
where $c_{ij} \equiv \cos\theta_{ij}$ and  $s_{ij} \equiv \sin\theta_{ij}$.
As a consequence, the Hamiltonian $H$  is a real  symmetric matrix 
\footnote{Replacing the matrix $U$ with $U^{sp} = UP^*_{33}(\delta)$
in eq. (\ref{eq:Hm}), it is easy to convince oneself that the Hamiltonian 
$H$ has the form given in eq. (\ref{eq:Hm3}) also in the standard 
parametrisation of the PMNS matrix with $U_{e3}$ replaced by 
$|U_{e3}| = s_{13}$.
}.
This implies that the matrix $U^\dagger U^m$, which diagonalises
$H$, is a real orthogonal matrix:
%%%%%%%%%%%%%%%%%%%%%%%%%
\begin{equation}
U^\dagger U^m = O\,,~~O^\ast = O\,,~~O^T\,O = O\,O^T = {\rm diag}(1,1,1)\,.
\label{eq:O}
\end{equation}
%%%%%%%%%%%%%%%%%%%%
%
Since $H$  does not depend on 
$\theta_{23}$ and $\delta$, $O=U^\dagger U^m$ 
should not depend on $\theta_{23}$ and $\delta$ either.
The fact that the matrix $O$ in eq. (\ref{eq:O})
is a real orthogonal matrix implies that in the parametrisations 
(\ref{eq:PMNSKP1}) and (\ref{eq:PMNSKP1m}) of the PMNS matrix 
in vacuum and in matter, the matrix 
%%%%%%%%%%%%%%%%%%%%%%%%%
\begin{align}
\tilde{O} & =  R_{13}(\theta_{13})\,R_{12}(\theta_{12})\,O\,R^T_{12}(\theta^m_{12})
\,R^T_{13}(\theta^m_{13})=P^\ast_{33}(\delta)\,R^T_{23}(\theta_{23})\,Q\,
\,R_{23}(\theta^m_{23})\,P_{33}(\delta^m)\\
& = 
\begin{pmatrix}
1 & 0 & 0\\
0 & c^m_{23}c_{23}e^{i\beta_2} +s^m_{23}s_{23}e^{i\beta_3} 
 & s^m_{23}c_{23}\,e^{i(\beta_2 +\delta^m)} - c^m_{23}s_{23}\,e^{i(\beta_3 +\delta^m)}\\
0 & s_{23}c^m_{23}\,e^{i(\beta_2 - \delta)} - c_{23}s^m_{23}\,e^{i(\beta_3 -\delta)} 
& s^m_{23}s_{23}e^{i(\beta_2-\delta +\delta^m)} +c^m_{23}c_{23}e^{i(\beta_3-\delta + \delta^m)} 
\end{pmatrix}\,,
\label{eq:tildeO}
\end{align}
%%%%%%%%%%%%%%%%%%%%
%
is a real orthogonal matrix.

The requirement of reality of 
the nondiagonal elements of $\tilde{O}$ 
leads to the conditions:
%%%%%%%%%%%%%%%%%%%%%
\begin{align}
\nonumber
& \cos\theta_{23}\sin\theta^m_{23}\,\sin(\beta_3 - \delta)  
= \sin\theta_{23}\cos\theta^m_{23}\,\sin(\beta_2 - \delta)\,,\\
& \cos\theta_{23}\sin\theta^m_{23}\,\sin(\beta_2 + \delta^m) 
= \sin\theta_{23}\cos\theta^m_{23}\,\sin(\beta_3 + \delta^m)\,,
\label{eq:tildeOreal1}
\end{align}
%%%%%%%%%%%%%%%%%%%%%%%%%
%
which imply, in particular:
%%%%%%%%%%%%%%%%%%%%%%%%%%%%%%
\begin{equation}
\cos(2\,\beta_3 + \delta^m - \delta) = \cos(2\,\beta_2 + \delta^m - \delta)\,.
\label{eq:tildeOreal2}
\end{equation}
%%%%%%%%%%%%%%%%%%%%%%%%%%%%%%%%%
%
The last condition has two solutions:
%%%%%%%%%%%%%%%%%%%%%%%%%%%
\begin{align}
\label{eq:tildeOreal3}
& \beta_3 = \beta_2 + k\,\pi\,,~~~k=0,1,2,...\,,\\
& \beta_2 + \beta_3  = \delta - \delta^m + k^\prime\pi\,,~~k^\prime=0,1,2,...\,.
\label{eq:tildeOreal4}
\end{align}
%%%%%%%%%%%%%%%%%%%%%%%%%%%%%%%%
%
The requirement of reality of 
the diagonal elements of $\tilde{O}$ 
leads to:
% the conditions:
%%%%%%%%%%%%%%%%%%%%%
\begin{align}
\nonumber
& \cos\theta_{23}\cos\theta^m_{23}\,\sin\beta_2  
= -\,\sin\theta_{23}\sin\theta^m_{23}\,\sin\beta_3\,,\\
& \cos\theta_{23}\cos\theta^m_{23}\,\sin(\beta_3 -\delta + \delta^m) 
= -\, \sin\theta_{23}\sin\theta^m_{23}\,\sin(\beta_2 -\delta + \delta^m)\,.
\label{eq:tildeOreal5}
\end{align}
%%%%%%%%%%%%%%%%%%%%%%%%%
%
These conditions also lead, in particular, to the constraint 
given in eq. (\ref{eq:tildeOreal2})
and to the solutions (\ref{eq:tildeOreal3}) and 
(\ref{eq:tildeOreal4}).
It should be clear that satisfying the constraint (\ref{eq:tildeOreal3})
or (\ref{eq:tildeOreal4}) is not enough to ensure the reality 
of the matrix  $\tilde{O}$. 

 Consider first the consequences of the constraint 
in eq. (\ref{eq:tildeOreal3}).
Requiring in addition that the determinant of  $\tilde{O}$ is 
a real quantity implies:
%%%%%%%%%%%%%%%%%%%%%%%%%%%%%%%%
\begin{equation}
2\beta_{2} = \delta - \delta^m + k^\prime\pi\,,~~~k^\prime=0,1,2,...
\label{eq:2b2}
\end{equation}
%%%%%%%%%%%%%%%%%%%%%%%%%%%%%%%%
%
The constraint in eq. (\ref{eq:tildeOreal3}) for $k=0$, for example,  
and the conditions of reality of the elements
$(\tilde{O})_{23(32)}$ and $(\tilde{O})_{22(33)}$
of $\tilde{O}$ lead to:
%%%%%%%%%%%%%%%%%%%%%%%%%%%%%%%%%%%%
\begin{equation}
\sin(\theta_{23} - \theta^m_{23})\,\sin(\beta_2 - \delta) = 0\,,~~
\sin(\theta_{23} - \theta^m_{23})\,\sin(\beta_2 + \delta^m) = 0\,.
\label{eq:tildeOreal32}
\end{equation}
%%%%%%%%%%%%%%%%%%%%%%%%%%%%
\begin{equation}
\cos(\theta_{23} - \theta^m_{23})\,\sin\beta_2 = 0\,,~~
\cos(\theta_{23} - \theta^m_{23})\,\sin(\beta_2 -\delta + \delta^m) = 0\,.
\label{eq:tildeOreal2233}
\end{equation}
%%%%%%%%%%%%%%%%%%%%%%%%%%%%
% 
As a consequence of eq. (\ref{eq:2b2}) 
the second conditions in eqs. (\ref{eq:tildeOreal32}) and 
(\ref{eq:tildeOreal2233}) are equivalent to the 
first conditions in eqs. (\ref{eq:tildeOreal32}) and 
(\ref{eq:tildeOreal2233}).
The constraint  (\ref{eq:tildeOreal3}) for $k =0$
and the conditions (\ref{eq:tildeOreal32}) and 
(\ref{eq:tildeOreal2233}) can be simultaneously 
satisfied if the following relations hold 
\footnote{An alternative solution to the discussed 
constraints is $\theta^m_{23} \neq \theta_{23}$,
$\beta_2 = q\pi$, $q=0,1,2,...$, 
$\delta = k^\prime\pi$, $k^\prime = 0,1,2$, 
$\delta^m = \tilde{k}^\prime\pi$, $\tilde{k}^\prime = 0,1,2$.
It corresponds to CP (T) conserving values of 
$\delta$ ($\delta^m$).
}:
%%%%%%%%%%%%%%%%%%%%%%%%
\begin{align} 
\label{eq:thm23eqth23}
& \theta^m_{23} = \theta_{23}\,,~~~~0 < \theta_{23},\theta^m_{23}\leq \pi/2\,,\\
\label{eq:dmd}
& \delta^m = \delta + k^\prime\pi\,,~~~ k^\prime = 0,1,2\,,\\
& \beta_2 = q\pi\,,~~~ q=0,1,2,...\,.
\label{eq:beta2qpi}
\end{align}
%%%%%%%%%%%%%%%%%%%%%%%%%%%
%
Numerical results on the dependence of 
$\theta^m_{23}$, $\delta^m$,  
$\overline{\theta}{}^m_{23}$ and $\overline{\delta}{}^m$
on the matter potential $A$,
$\overline{\theta}{}^m_{23}$ and $\overline{\delta}{}^m$
being the corresponding antineutrino mixing angle and CP 
violation phase,
show that the relations (\ref{eq:thm23eqth23}) and (\ref{eq:dmd}) 
cannot be exact. These relations are not the true solutions of the reality conditions of
the matrix  $\tilde{O}$ since they do not fully
guarantee the reality of  $\tilde{O}$ as given in eq. (27).
Only if both $\theta_{23} = 45^\circ$ and $\delta = \pm \pi/2$ hold in vacuum,
the relations (36) and (37) are exact and are not violated by the
effects of matter \cite{Xing:2010xy}. However, they are fulfilled with extremely 
high precision for the mixing of neutrinos (antineutrinos) in matter 
in the case of IO (NO) neutrino mass spectrum.
We find that in this case  for any $A/\Delta m^2_{21}$ 
and the best fit values of the neutrino oscillation parameters 
quoted in eq. (\ref{eq:bfvnoscparam}) we have:
%%%%%%%%%%%%%%%%%%%
\begin{align}
\label{eq:thm23th23a}
&\bigg| \dfrac{\overset{\scriptscriptstyle(-)}{\theta}\!{}^m_{23}}{\theta_{23}} - 1 \bigg| 
\lesssim 0.0004~(0.0015)\,,\\
&\bigg| \dfrac{\overset{\scriptscriptstyle(-)}{\delta}\!{}^m}{\delta} - 1 \bigg| 
\lesssim 0.0001~(0.00006)\,. 
\label{eq:dmda}
\end{align}
%%%%%%%%%%%%%%%%%%%%%%%%%%%%%
%
For the mixing of neutrinos (antineutrinos) in matter 
and spectrum with normal (inverted) ordering,
eqs. (\ref{eq:thm23eqth23}) and (\ref{eq:dmd})
are  fulfilled also with extremely high precision 
for $A/\Delta m^2_{21} < 30$:
%%%%%%%%%%%%%%%%%%%
\begin{align}
\label{eq:thm23th23b}
&\bigg| \dfrac{\overset{\scriptscriptstyle(-)}{\theta}\!{}^m_{23}}{\theta_{23}} - 1 \bigg| 
\lesssim 0.006~(0.0015)\,,\\
&\bigg| \dfrac{\overset{\scriptscriptstyle(-)}{\delta}\!{}^m}{\delta} - 1 \bigg| \lesssim 0.0003~(0.001)\,.
\label{eq:dmdb}
\end{align}
%%%%%%%%%%%%%%%%%%%%%%%%%%%%%
%
For 
\footnote{For an analytic understanding of the results in eqs. (41) and (42)  
see \cite{Ioannisian:2018qwl}.}
$A/\Delta m^2_{21} \gtrsim 30$ and 
mixing of neutrinos (antineutrinos) in matter 
and NO (IO) neutrino mass spectrum we have:
%%%%%%%%%%%%%%%%%%%
\begin{align}
\label{eq:thm23th23c}
&\bigg| \dfrac{\overset{\scriptscriptstyle(-)}{\theta}\!{}^m_{23}}{\theta_{23}} - 1 \bigg| \lesssim 0.07~(0.016)\,,\\
&\bigg| \dfrac{\overset{\scriptscriptstyle(-)}{\delta}\!{}^m}{\delta} - 1 \bigg| \lesssim 0.001~(0.004)\,.
\label{eq:dmdc}
\end{align}
%%%%%%%%%%%%%%%%%%%%%%%%%%%%%
%

Setting $\delta$ to its best fit value given in 
eq. (\ref{eq:bfvnoscparam}) and varying the other 
neutrino oscillation parameters in their 
$3\sigma$ allowed ranges determied in  \cite{Esteban:2016qun} 
does not change significantly the results quoted in 
eqs. (\ref{eq:dmda}) - (\ref{eq:dmdc}). 
Indeed,  for mixing of neutrinos (antineutrinos) in matter 
in the case of IO (NO) neutrino mass spectrum 
and any $A/\Delta m^2_{21}$ we find that 
$|\overset{\scriptscriptstyle(-)}{\theta}\!{}^m_{23}/\theta_{23} - 1| \lesssim 0.0005~(0.002)$ and
$|\overset{\scriptscriptstyle(-)}{\delta}\!{}^m/\delta - 1| \lesssim 0.0002~(0.0002)$.
In the case of mixing of neutrinos (antineutrinos) in matter and 
NO (IO) spectrum and  $A/\Delta m^2_{21} < 30$ we get
$|\overset{\scriptscriptstyle(-)}{\theta}\!{}^m_{23}/\theta_{23} - 1| \lesssim 0.013~(0.003)$ and
$|\overset{\scriptscriptstyle(-)}{\delta}\!{}^m/\delta - 1| \lesssim 0.0013~(0.003)$,
while for $A/\Delta m^2_{21} \gtrsim 30$ we obtain 
$|\overset{\scriptscriptstyle(-)}{\theta}\!{}^m_{23}/\theta_{23} - 1| \lesssim 0.09~(0.02)$ and
 $|\overset{\scriptscriptstyle(-)}{\delta}\!{}^m/\delta - 1|\lesssim 0.01~(0.01)$. \\[-1mm]

These results are illustrated in Figs. \ref{Fig3} and   \ref{Fig32}
(\ref{Fig33} and   \ref{Fig34})
where the ratios  $\theta^m_{23}/\theta_{23}$ 
and $\delta^m/\delta$ 
(the ratios  $\overline{\theta}{}^m_{23}/\theta_{23}$ 
and  $\overline{\delta}{}^m/\delta$)
are shown as  functions of $A/\Delta m^2_{21}$
in the case of mixing of neutrinos 
(antineutrinos) 
and NO (left panel) and IO (right panel) 
neutrino mass spectrum.
We used  the best fit values of neutrino oscillation parameters
$\Delta m^2_{31}$,  $\Delta m^2_{21}$, $\theta_{12}$ and $\theta_{13}$
from \cite{Esteban:2016qun} and 
the analytic expressions for 
$M^2_i$, $i=1,2,3$, 
from \cite{Zaglauer:1988gz}.
%%%%%%%%%%%%%%%%%%%%%%%%%%%%%%%%%%%%%%
\begin{figure}[t]
\vspace{-0.5cm}
\centering
\includegraphics[scale=0.7]{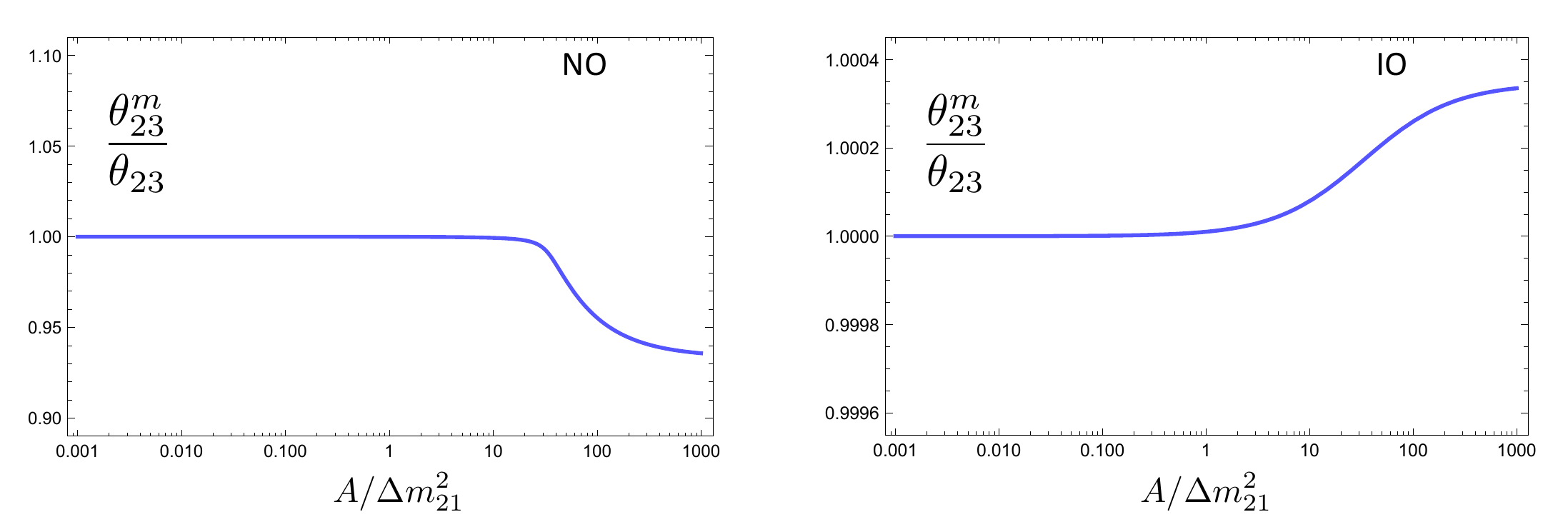}
\vskip -0.4cm
\caption{ The ratio
$\theta^m_{23}/\theta_{23}$ 
as a function of $A/\Delta m^2_{21}$
in the case of mixing of neutrinos 
and NO (left panel) and IO (right panel) 
neutrino mass spectrum.
See text for further details.
}
\label{Fig3}
\vspace{-0.1cm}
\end{figure}
%%%%%%%%%%%%%%%%%%%%%%%%%%%%%%%%%%%%%%%%%%%%%%%%
%

%%%%%%%%%%%%%%%%%%%%%%%%%%%%%%%%%%%%%%
\begin{figure}[htb]
\vspace{-0.1cm}
\centering
\includegraphics[scale=0.7]{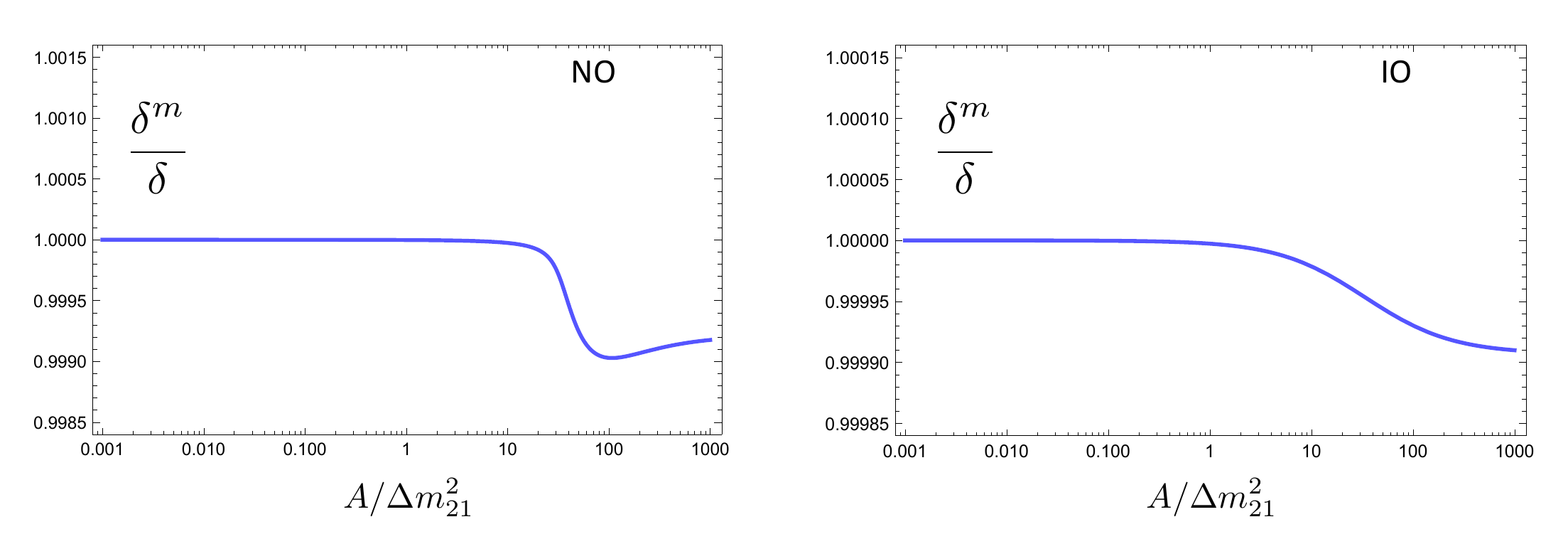}
\vskip -0.4cm
\caption{The ratio 
$\delta^m/\delta$ 
as a function of $A/\Delta m^2_{21}$ 
in the case of mixing of neutrinos 
and NO (left panel) and IO (right panel) 
neutrino mass spectrum.
See text for further details.
}
\label{Fig32}
\vspace{-0.1cm}
\end{figure}
%%%%%%%%%%%%%%%%%%%%%%%%%%%%%%%%%%%%%%%%%%%%%%%%
%

The approximate ranges of values of $A/\Delta m^2_{21}$ 
relevant for the T2K (T2HK) \cite{T2K20152016}, T2HKK \cite{T2HKK2016},
NO$\nu$A \cite{NOvA2016} and DUNE \cite{DUNE2016}
long baseline neutrino oscillation experiments read, respectively:
[0.266,2.66], [0.306,3.06], [2.90,8.70] and [3.02,12.10].}
In obtaining these ranges 
we used the best fit value of 
$\Delta m^2_{21} = 7.4\times 10^{-5}~{\rm eV^2}$ and
took into account  
i) that $A = 7.56\times 10^{-5}~{\rm eV^2}\,({\rm \rho/g/cm^3})\,(E/{\rm GeV})$,
where $\rho$ is the matter density, 
ii) that the mean Earth density 
along the trajectories of the neutrinos in the 
T2K (T2HK), T2HKK, NO$\nu$A and DUNE 
long baseline neutrino oscillation 
experiments respectively is 2.60, 3.00, 2.84 and 2.96 g/cm$^3$, 
and iii) that in these experiments beams of neutrinos with energies 
$\sim (0.1 - 1.0)$ GeV (T2K, T2HK, T2HKK), $\sim (1 - 3)$ GeV (NO$\nu$A) and 
$\sim (1 - 4)$ GeV (DUNE) are being, or planned to be, used. 
At the peak neutrino energies at T2K (T2HK), T2HKK, NO$\nu$A and DUNE 
experiments 
of respectively 0.6 GeV, 0.6 GeV, 2.0 GeV and 2.6 GeV we have 
 $A/\Delta m^2_{21} = 1.59$,~1.84,~5.80 and 7.86.
Taking a wider neutrino energy interval for, e.g.,  
NO$\nu$A and DUNE experiments of [1.0,8.0] GeV, 
we get for the corresponding $A/\Delta m^2_{21}$ ranges:
[2.90,23.21] and [3.02,24.20].
For all the intervals of values of $A/\Delta m^2_{21}$ 
quoted above, which are relevent for the 
T2K (T2HK), T2HKK, NO$\nu$A and DUNE experiments,
the equalities (\ref{eq:thm23eqth23}) 
and (\ref{eq:dmd}) 
are excellent approximations.

  Consider next the implications of the 
second condition (\ref{eq:tildeOreal4}) 
related to the requirement of reality of the matrix $\tilde{O}$.
As can be easily shown, this condition alone
i) ensures the reality of  ${\rm det}(\tilde{O})$,
and ii) makes identical the two conditions in eq. (\ref{eq:tildeOreal1})
and the two conditions in eq. (\ref{eq:tildeOreal5}).
Thus, after using condition  (\ref{eq:tildeOreal4}) 
there are still two independent conditions to be satisfied to 
ensure the reality of the matrix  $\tilde{O}$. We will derive next a condition 
that can substitute one of the required two conditions. 
The second condition then can be either the condition in 
eq.  (\ref{eq:tildeOreal1}) or the condition in eq. (\ref{eq:tildeOreal5})
(after eq. (\ref{eq:tildeOreal4}) has been used). 
%%%%%%%%%%%%%%%%%%%%%%%%%%%%%%%%%%%%%%
\begin{figure}[t]
\vspace{-0.5cm}
\centering
\includegraphics[scale=0.7]{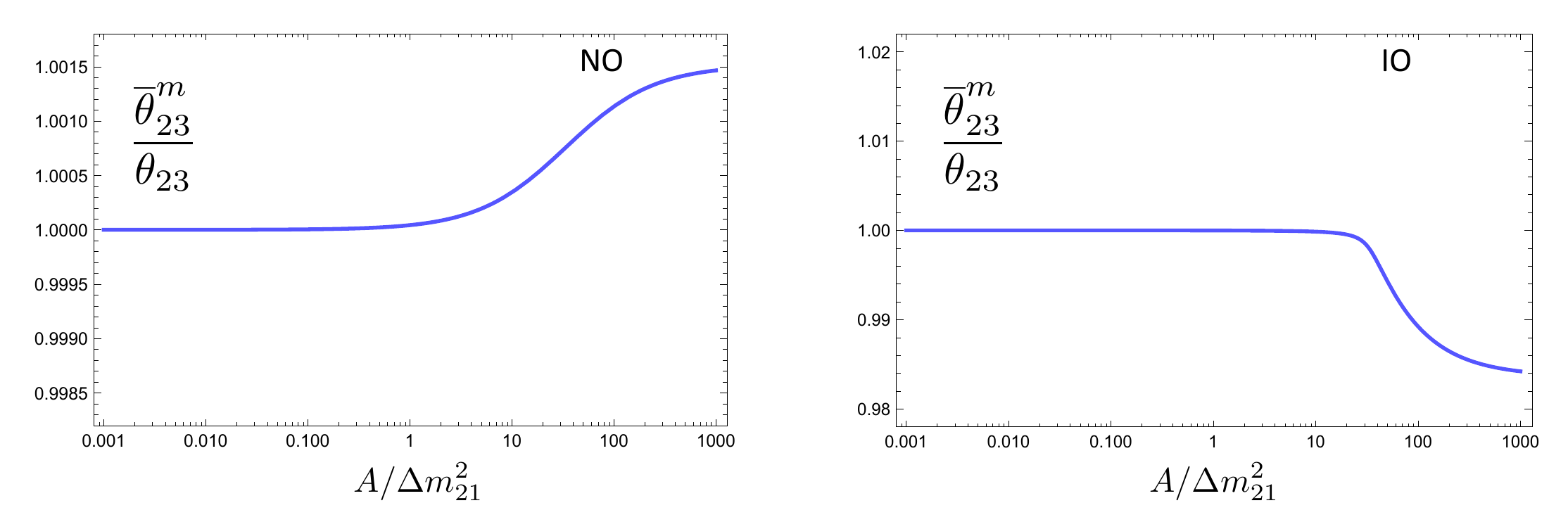}
\vskip -0.4cm
\caption{ The same as in Fig. \ref{Fig3} but for the ratio $\overline{\theta}{}^m_{23}/\theta_{23}$.
See text for further details. }
\label{Fig33}
\vspace{-0.1cm}
\end{figure}
%%%%%%%%%%%%%%%%%%%%%%%%%%%%%%%%%%%%%%%%%%%%%%%%
%

%%%%%%%%%%%%%%%%%%%%%%%%%%%%%%%%%%%%%%
\begin{figure}[htb]
\vspace{-0.1cm}
\centering
\includegraphics[scale=0.7]{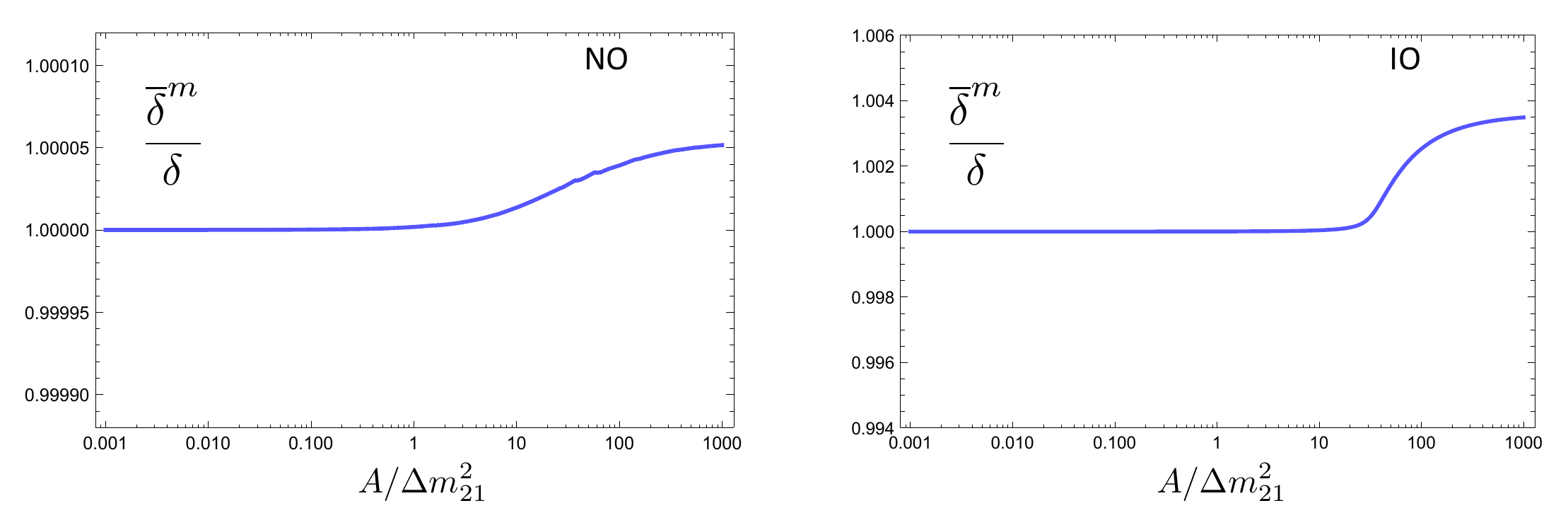}
\vskip -0.4cm
\caption{The same as Fig. \ref{Fig32} 
but for the ratio $\overline{\delta}{}^m/\delta$.
See text for further details.
}
\label{Fig34}
\vspace{-0.1cm}
\end{figure}
%%%%%%%%%%%%%%%%%%%%%%%%%%%%%%%%%%%%%%%%%%%%%%%%
%

 The condition of orthogonality of 
$\tilde{O}$,   $\tilde{O}(\tilde{O})^T = {\rm diag}(1,1,1)$, as can 
be shown, leads to the following additional constraints:
%%%%%%%%%%%%%%%%%%%%%%%%%%% 
\begin{align}
\label{eq:tOtOt23}
(c^m_{23})^2\,\sin(2\beta_2 - \delta) + 
(s^m_{23})^2\,\sin(2\beta_2 - \delta + 2\delta^m) = 
-\, \dfrac{c^m_{23}s^m_{23}}{c_{23}s_{23}}\,\cos2\theta_{23}\,\sin\delta^m\,,\\  
\nonumber
\sin2\theta_{23}\,\sin2\theta^m_{23}\,\sin\delta\,\sin\delta^m
+ (c^m_{23})^2\,\cos(2\beta_2) + 
(s^m_{23})^2\,\cos(2\beta_2 + 2\delta^m)\\ 
\label{eq:RetOtOt22}
= 1\,-\,2\,s^2_{23}\sin\delta\,[(c^m_{23})^2\,\sin(2\beta_2 - \delta) + 
(s^m_{23})^2\,\sin(2\beta_2 - \delta + 2\delta^m)]\,,\\
\nonumber
-\,\sin2\theta_{23}\,\sin2\theta^m_{23}\,\cos\delta\,\sin\delta^m
+ (c^m_{23})^2\,\sin(2\beta_2) + (s^m_{23})^2\,\sin(2\beta_2 + 2\delta^m)\\
= 2\,s^2_{23}\cos\delta\,[(c^m_{23})^2\,\sin(2\beta_2 - \delta) + 
(s^m_{23})^2\,\sin(2\beta_2 - \delta + 2\delta^m)]\,,
\label{eq:ImtOtOt22}
\end{align}
%%%%%%%%%%%%%%%%%%%%%%%%%%%%%%%%
%
where we have used the relation in eq. (\ref{eq:tildeOreal4}).
Conditions (\ref{eq:tOtOt23}), (\ref{eq:RetOtOt22}) and 
(\ref{eq:ImtOtOt22}) follow from the requirements 
 $(\tilde{O}(\tilde{O})^T)_{23(32)} = 0$,
${\rm Re}((\tilde{O}(\tilde{O})^T)_{22}) = 1$ and 
${\rm Im}((\tilde{O}(\tilde{O})^T)_{22}) = 0$, respectively.
Replacing $(c^m_{23})^2\,\sin(2\beta_2 - \delta) + 
(s^m_{23})^2\,\sin(2\beta_2 - \delta + 2\delta^m)$ in 
eqs.  (\ref{eq:RetOtOt22}) and (\ref{eq:ImtOtOt22}) with 
the right hand side of eq. (\ref{eq:tOtOt23}), 
after certain simple algebra leads to the equality:
%%%%%%%%%%%%%%%%%%%%%%%%%%%%%%%%%
\begin{equation}
\sin2\theta^m_{23}\,\sin\delta^m
= \sin2\theta_{23}\,\sin\delta\,.
\label{eq:s2th23mdm}
\end{equation}
%%%%%%%%%%%%%%%%%%%%%%%%%%%%%%%
%
This result was derived in \cite{Toshev:1991ku} 
(see also \cite{Freund:2001pn}) using the 
parametisations (\ref{eq:PMNSKP1})
and (\ref{eq:PMNSKP1m}) introduced 
in \cite{Krastev:1988yu} but employing a different method 
\footnote{The method employed in \cite{Toshev:1991ku} 
is based on the observation \cite{Petcov:1987cd} that 
the parametrisation (\ref{eq:PMNSKP1}) allows to factor out 
the part  $R_{23}(\theta_{23})P_{33}(\delta)$
in the neutrino mixing matrix in matter.
In this case one works with the Hamiltonian 
$\hat{H} = P^\ast_{33}(\delta)R^T_{23}(\theta_{23})UHU^\dagger
R_{23}(\theta_{23})P_{33}$, 
which is also a real symmetric matrix, 
where $H$ is given in eq. (\ref{eq:Hm3}).
}. 
The equality  (\ref{eq:s2th23mdm}) 
implies that the product
$\sin2\theta_{23}\sin\delta$ does not depend on the matter 
potential, i.e., is the same for neutrino oscillations taking 
place in vacuum and in matter with constant density. 
%%%%%%%%%%%%%%%%%%%%%%%%%%%%%%%%%%%%%%
\begin{figure}[t]
\vspace{-0.5cm}
\centering
\includegraphics[scale=0.7]{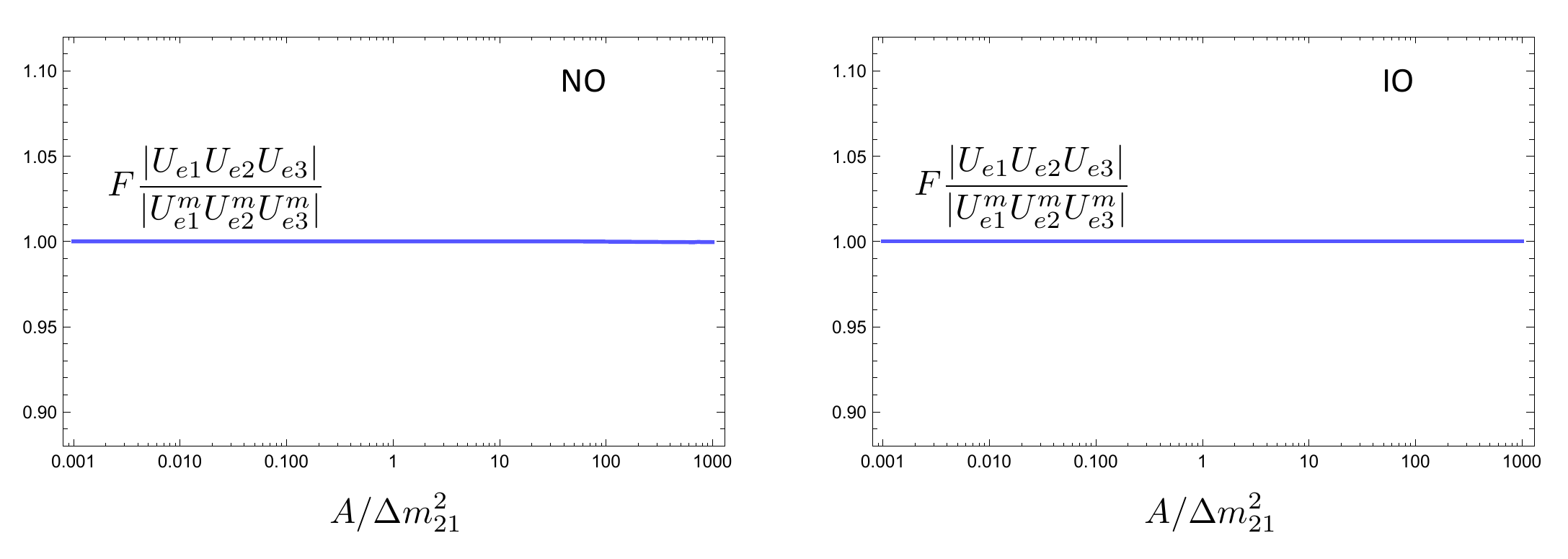}
\vskip -0.4cm
\caption{The ratio 
$\sin2\theta^m_{23}\sin\delta^m/
(\sin2\theta_{23}\sin\delta)$, eq. (\ref{eq:thm23dmth23d}),  
for mixing of neutrinos 
as a function of $A/\Delta m^2_{21}$ 
in the cases of NO (left panel) and 
IO (right panel) neutrino mass spectra.
See text for further details.
}
\label{Fig2}
\vspace{-0.1cm}
\end{figure}
%%%%%%%%%%%%%%%%%%%%%%%%%%%%%%%%%%%%%%%%%%%%%%%%
%
\noindent It is valid for neutrino and antineutrino mixing in matter 
independently of the type of spectrum 
neutrino masses obey - with NO
or IO.
From eq. (\ref{eq:JCPmJCPv}) using the parametrisations 
defined in eqs. (\ref{eq:PMNSKP1}) and 
(\ref{eq:PMNSKP1m}) we find: 
%%%%%%%%%%%%%%%%%%%%%%%%%%%
\begin{equation}
\dfrac{\sin2\theta^m_{23}\sin\delta^m}
{\sin2\theta_{23}\sin\delta} =
F\,\dfrac{\cos\theta_{13}\,\sin 2\theta_{12}\,\sin 2\theta_{13}}
{\cos\theta^m_{13}\,\sin 2\theta^m_{12}\,\sin 2\theta^m_{13}}
= F\,\dfrac{U_{e1}\,U_{e2}\,U_{e3}}
{U^m_{e1}\,U^m_{e2}\,U^m_{e3}}\,.    
\label{eq:thm23dmth23d}
\end{equation}
%%%%%%%%%%%%%%%%%%%%%%%%%%%%%%%%%%%
%
From this result and eq. (\ref{eq:s2th23mdm})  
we obtain yet another equivalent 
representation of the function 
$F(\theta_{12},\theta_{13},\Delta m^2_{21},\Delta m^2_{31},A)$:   
%%%%%%%%%%%%%%%%%%%%%%%%%%%%%%%%%
\begin{equation}
F = \dfrac{U^m_{e1}\,U^m_{e2}\,U^m_{e3}}{U_{e1}\,U_{e2}\,U_{e3}}\,.
\label{eq:FUejUmej}
\end{equation}
%%%%%%%%%%%%%%%%%%%%%%%%%%%%%%
%

The ratio given in eq. (\ref{eq:thm23dmth23d})
is shown graphically in Fig. \ref{Fig2}
for mixing of neutrinos as a function 
of  $A/\Delta m^2_{21}$ for the best fit values 
of neutrino oscillation parameters
$\Delta m^2_{31}$,  $\Delta m^2_{21}$, $\theta_{12}$ and $\theta_{13}$
from \cite{Esteban:2016qun} and the analytic expressions for 
$M^2_i$, $i=1,2,3$, 
from \cite{Zaglauer:1988gz}. 
The numerical result presented in Fig. \ref{Fig2}, 
as we have verified and could be expected, 
is valid not only for best fit
values of the relevant neutrino oscillation parameters, 
but indeed holds for any values of these parameters, 
varied in their respective physical regions. 
The same result is valid for mixing of antineutrinos.
Thus, the equality (\ref{eq:s2th23mdm}) is exact. It 
holds also in the standard parametrisations of the 
PMNS matrix (see footnote 4). 
In this case the ratio $U_{e3}/U^m_{e3}$ in 
eqs. (\ref{eq:thm23dmth23d}) and (\ref{eq:FUejUmej}) 
has to be replaced by  $|U_{e3}|/|U^m_{e3}|$.

  It follows from eq. (\ref{eq:s2th23mdm})
that for the values of $\sin2\theta_{23} = 1$ and $\delta = 3\pi/2$, 
which are perfectly compatible with the existing data,
we have $\sin2\theta^m_{23}\,\sin\delta^m = -\,1$. 
This in turn implies $\sin2\theta^m_{23} = 1$ and $\sin\delta^m = -\,1$, 
i.e., the vacuum values of 
$\theta_{23} = \pi/4$ and  $\delta = 3\pi/2$ 
are not modified by the presence of matter \cite{Xing:2010xy}.

Given the fact that, as we have seen, the corrections of
$\theta_{23}$ and $\delta$ due to the matter effects are small,
the relation  (\ref{eq:s2th23mdm}) allows to relate 
the matter correction to $\theta_{23}$, $\epsilon_{23}(A/\Delta m^2_{21})$,
with the matter correction to $\delta$, 
$\epsilon_{\delta}(A/\Delta m^2_{21})$.
Working to leading order in  $\epsilon_{23}\ll 1$ and  
$\epsilon_{\delta}\ll 1$ 
we get from eq. (\ref{eq:s2th23mdm}) using  
$\theta^m_{23} = \theta_{23} + \epsilon_{23}$ and 
$\delta^m = \delta + \epsilon_{\delta}$:
%%%%%%%%%%%%%%%%%%%%%%%%%%
\begin{equation}
\epsilon_{\delta}(A/\Delta m^2_{21})\cos\delta \cong 
-\,2\,\epsilon_{23}(A/\Delta m^2_{21})\,
\dfrac{\cos2\theta_{23}}{\sin2\theta_{23}}\, \sin\delta\,.
\label{eq:corrth23d} 
\end{equation}
%%%%%%%%%%%%%%%%%%%%%%%%%%%%%%%
%
Thus, for $\theta_{23} = \pi/4$ 
and $\delta \neq 3\pi/2,\pi/2$, 
the leading order matter 
correction to $\delta$ vanishes, while for 
$\delta = 3\pi/2$ ($\pi/2$) and 
$\theta_{23}\neq \pi/4$, the leading order matter correction 
to $\theta_{23}$ vanishes.
For $\delta \neq q\pi/2$, $q=0,1,2,3,4$, and 
$\theta_{23}\neq \pi/4$ the sign of $\epsilon_{\delta}$
coincides with (is opposite to) the sign of $\epsilon_{23}$
provided $\cos2\theta_{23}\cot\delta < 0$ 
($\cos2\theta_{23}\cot\delta > 0$).

%%%%%%%%%%%%%%%%%%%%%%%%%%%%%%%%%%
%
\section{The Relation between $J^m$ and $J_{\rm CP}$} 
%
%%%%%%%%%%%%%%%%%%%%%%%%%%%%%
 
 Equation (\ref{eq:Hm}) can be cast in the form:  
%%%%%%%%%%%%%%%%%%%%%%%%%%% 
\begin{equation}
\left [ 
\begin{pmatrix}
m^2_1 & 0 & 0\\
0 & m^2_2 & 0\\
0 & 0 & m^2_3 
\end{pmatrix}
+
U^\dagger\,
\begin{pmatrix}
A & 0 & 0\\
0 & 0 & 0\\
0 & 0 & 0 
\end{pmatrix}\,U\right ](U^\dagger\,U^m) = 
 (U^\dagger\,U^m)\,
\begin{pmatrix}
M^2_1 & 0 & 0\\
0 & M^2_2 & 0\\
0 & 0 & M^2_3 
\end{pmatrix}\,.
\label{eq:Hm2} 
\end{equation}
%%%%%%%%%%%%%%%%%%%%%%%%%%%
% 

 One possible relatively simple 
way to derive the relation 
between $J^m$ and $J_{\rm CP}$ 
given in eq. (\ref{eq:JCPmJCPv2})
and reported in \cite{Krastev:1988yu}
is to exploit the fact that the column matrices\\  
\noindent 
$((U^\dagger U^m)_{1i}~(U^\dagger U^m)_{2 i}~(U^\dagger U^m)_{3 i})^T$ 
are eigenvectors of the Hamiltonian  
$H$ defined in eq. (\ref{eq:Hm3}),
corresponding to the eigenvalues $M^2_i/(2E)$, $i=1,2,3$.
Using this observation it is possible to 
derive from eq. (\ref{eq:Hm2}) explicit expressions 
for the elements of the neutrino mixing matrix 
in matter $U^m$. They read:
%%%%%%%%%%%%%%%%%%%%%%%%%%%%%%%%
\begin{equation}
U^m_{li} = \frac{1}{D_i}\,\left[N_i\,U_{li} - 
A\,U_{ei}\left (D_{ji}\,U^*_{ek}\,U_{lk} + D_{ki}\,U^*_{ej}\,U_{lj}\right )\right ]\,,~l=e,\mu,\tau\,,
\label{eq:Umli}
\end{equation}
%%%%%%%%%%%%%%%%%%%%%%%%%%%%%%%
%
where 
%%%%%%%%%%%%%%%%%%%%%%%
\begin{eqnarray}
\label{eq:Ni}
& N_i = D_{ji}\,D_{ki} + A\left (D_{ji}\,|U_{ek}|^2 + D_{ki}\,|U_{ej}|^2\right)\,,
\\[0.3cm]
& D^2_i = N^2_i + A^2\,|U_{ei}|^2\left (D^2_{ji}\,|U_{ek}|^2 
+ D^2_{ki}\,|U_{ej}|^2\right)\,,
\label{eq:Disqr}
\end{eqnarray}
%%%%%%%%%%%%%%%%%%%%%%%%%%%%%%%
%
with $i,j,k=1,2,3$, but $i \neq j \neq k\neq i$.
For the elements of $U^m$ of interest, $U^m_{e2}$, $U^m_{e3}$, $U^m_{\mu 2}$ 
and  $U^m_{\mu 3}$ we get from eqs. (\ref{eq:Umli}) - (\ref{eq:Disqr}):
%%%%%%%%%%%%%%%%%%%%%%%
\begin{equation}
U^m_{e2} = \frac{1}{D_2}\,U_{e2}\,D_{12}\,D_{32}\,,
\label{eq:Ume2}
\end{equation}
%%%%%%%%%%%%%%%%%%%%%%%%%%%%%%%
%
%%%%%%%%%%%%%%%%%%%%%%%
\begin{equation}
U^m_{e3} = \frac{1}{D_3}\,U_{e3}\,D_{13}\,D_{23}\,,
\label{eq:Ume3}
\end{equation}
%%%%%%%%%%%%%%%%%%%%%%%%%%%%%%%
%
%%%%%%%%%%%%%%%%%%%%%%%
\begin{equation}
U^m_{\mu 2} = \frac{1}{D_2}\,\left[N_2\,U_{\mu 2} - 
A\,U_{e2}\left (D_{12}\,U^*_{e3}\,U_{\mu 3} 
+ D_{32}\,U^*_{e1}\,U_{\mu 1}\right )\right ]\,,
\label{eq:Ummu2}
\end{equation}
%%%%%%%%%%%%%%%%%%%%%%%%%%%%%%%
%
and
%%%%%%%%%%%%%%%%%%%%%%%
\begin{equation}
U^m_{\mu 3} = \frac{1}{D_3}\,\left[N_3\,U_{\mu 3} - 
A\,U_{e3}\left (D_{13}\,U^*_{e2}\,U_{\mu 2} + D_{23}\,U^*_{e1}\,U_{\mu 1}\right )\right ]\,.
\label{eq:Ummu3}
\end{equation}
%%%%%%%%%%%%%%%%%%%%%%%%%%%%%%%
%
The function $D^2_2$, which, as it follows from 
eqs. (\ref{JCPvm}) and (\ref{eq:Ume2}) -  (\ref{eq:Ummu3}),
enters into the expression for 
$J^m$, is given by:
%%%%%%%%%%%%%%%%%%%%%%%%%%%%%%%%%%%
\begin{eqnarray}
\label{eq:D2sqr1}
\nonumber 
& D^2_2 = N^2_2 + A^2\,|U_{e2}|^2\left( D^2_{12}\,|U_{e3}|^2 
+ D^2_{32}\,|U_{e1}|^2 \right) \\ [0.3cm]
\nonumber
& =\,D^2_{12}\,D^2_{32} 
+ 2\,A\,D_{12}\,D_{32}\,\left (D_{12}\,|U_{e3}|^2 
+ D_{32}\,|U_{e1}|^2\right) 
\\ [0.3cm]
\nonumber
& + A^2\,\left[D^2_{12}\,|U_{e3}|^2\left(|U_{e3}|^2+|U_{e2}|^2\right)
+ D^2_{32}\,|U_{e1}|^2\left(|U_{e1}|^2+|U_{e2}|^2\right) 
+ 2\,D_{12}\,D_{32}|U_{e1}|^2\,|U_{e3}|^2\right ]\\[0.3cm]
\nonumber
& = D^2_{12}\,D^2_{32} + 2\,A\,D_{12}\,D_{32}\,\left (D_{12}\,|U_{e3}|^2 
+ D_{32}\,|U_{e1}|^2\right)\\ [0.3cm]
& + A^2\left[ D^2_{12}\,|U_{e3}|^2 + D^2_{32}\,|U_{e1}|^2 
- \left (D_{32} - D_{12}\right)^2\,|U_{e1}|^2\,|U_{e3}|^2 \right]\,.
\label{eq:D2sqr2}
\end{eqnarray}
%%%%%%%%%%%%%%%%%%%%%%%%%%%%%%%%
%         
It is easy to check that expression (\ref{eq:D2sqr2}) 
for the function $D^2_2$ coincides with expression 
(\ref{eq:F2}) for the function $F_2$, i.e., that 
we have
%%%%%%%%%%%%%%%%%%%%%%%%%%%%%%%
\begin{equation}
D^2_2 = F_2\,.
\label{eq:D2sqrF2}
\end{equation}
%%%%%%%%%%%%%%%%%%%%%%%%%%%%%%%%%%%
%
One can show in a similar way that
the function $D^2_3$ coincides with the function 
$F_3$ given in eq. (\ref{eq:F3}), i.e., that 
%%%%%%%%%%%%%%%%%%%%%%%%%%%%%%%
\begin{equation}
D^2_3 = F_3\,.
\label{eq:D3sqrF3}
\end{equation}
%%%%%%%%%%%%%%%%%%%%%%%%%%%%%%%%%%%
%

The calculation of the rephasing invariant in matter 
$J^m$ 
involves, in particular, the product 
$U^m_{\mu 3}(U^m_{\mu 2})^\ast (U^m_{e 3})^\ast U^m_{e 2}$
of elements of $U^m$. 
From eqs. (\ref{eq:Ume2}) -  (\ref{eq:Ummu3}) we have:
%%%%%%%%%%%%%%%%%%%%%%%%%%%%%%%%%%%
\begin{align}
\label{eq:Rdef}
& R \equiv \dfrac{D^2_2\,D^2_3}{D_{13}\,D_{23}\,D_{12}\,D_{32}}\,
\dfrac{{\rm Im}\big( (U^m_{\mu 2})^\ast \,U^m_{\mu 3}\,(U^m_{e 3})^\ast \,U^m_{e 2} \big)}{{\rm Im}\big(U^\ast_{\mu 2}U_{e2}U_{\mu 3}U^\ast_{e3}\big)}
\\
\nonumber
& = \dfrac{1}{{\rm Im}\big(U^\ast_{\mu 2}U_{e2}U_{\mu 3}U^\ast_{e3}\big)}\,
 {\rm Im}\big( 
\big[N^\ast_2\,U^\ast_{\mu 2}\,U_{e 2} - 
A\,|U_{e2}|^2\left (D_{12}\,U_{e3}\,U^\ast_{\mu 3} 
+ D_{32}\,U_{e1}\,U^\ast_{\mu 1}\right)\big] 
\\
& \times 
\,\big[N_3\,U_{\mu 3}\,U^\ast_{e 3} - 
A\,|U_{e3}|^2 \left(D_{13}\,U^\ast_{e2}\,U_{\mu 2} 
+ D_{23}\,U^\ast_{e 1}\,U_{\mu 1} \right)\big] \big)\,.
\label{eq:R1}
\end{align}
%%%%%%%%%%%%%%%%%%%%%%%%%%%%%%
%
Using the fact that 
%%%%%%%%%%%%%%%%%%%%%%%%%%%
\begin{equation}
J_{\rm CP} = {\rm Im}(U^\ast_{\mu 2}U_{e2}U_{\mu 3}U^\ast_{e3})
=  {\rm Im}(U^\ast_{\mu 3}U_{e3}U_{\mu 1}U^\ast_{e1}) 
= -\, {\rm Im}(U^\ast_{\mu 2}U_{e2}U_{\mu 1}U^\ast_{e1})\,,
\label{eq:JCPv}
\end{equation}
%%%%%%%%%%%%%%%%%%%%%%%%
%
%%%%%%%%%%%%%%%%%%%%%%%%%%%%%%%%%%%%%%
\begin{figure}[t]
\vspace{-0.5cm}
\centering
\includegraphics[scale=0.7]{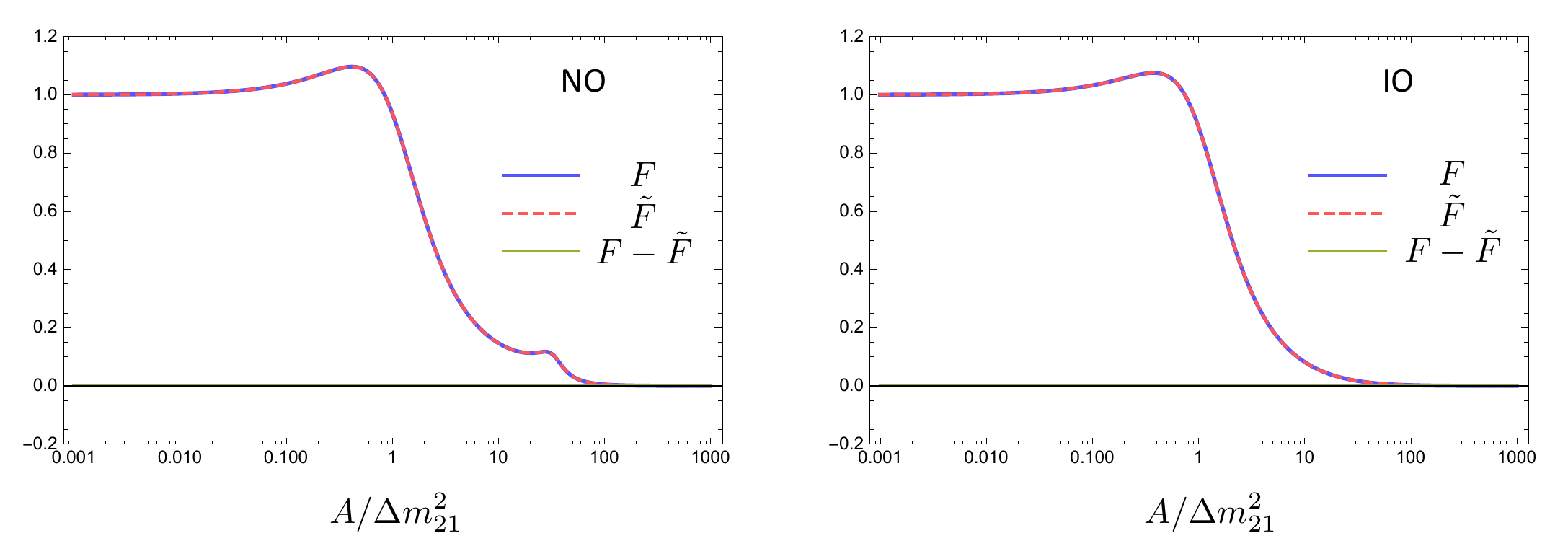}
\vskip -0.4cm
\caption{The functions $F$ (blue solid line), 
$\tilde{F}$ (red dashed line) 
and the difference $F - \tilde{F}$ (green line)
versus $A/\Delta m^2_{21}$ for NO (left panel) and 
IO (right panel) neutrino mass spectrum.
See text for further details.
}
\label{Fig1}
\vspace{-0.1cm}
\end{figure}
%%%%%%%%%%%%%%%%%%%%%%%%%%%%%%%%%%%%%%%%%%%%%%%%
%
\noindent
the function $R$ in eq. (\ref{eq:R1}), after some algebra, 
can be brought to the form:
%%%%%%%%%%%%%%%%%%%%%%%%%%%%%%%%%%%
\begin{align}
\nonumber
& R = D_{13}\,D_{23}\,D_{12}\,D_{32} + 
A\,\big[ D_{13}\,D_{23}\,\big (D_{32} 
- |U_{e3}|^2\left (D_{32}-D_{12}\right)\big)
+ D_{12}\,D_{32}\,\big(D_{23} 
\\
& - |U_{e2}|^2\left (D_{23}-D_{13}\right)\big)\big]
+ A^2\, \big[ D_{23}\,D_{32}\,|U_{e1}|^2 
+ D_{13}\,D_{32}\,|U_{e2}|^2 + D_{12}\,D_{23}\,|U_{e3}|^2 \big]\,.
\label{eq:R2}
\end{align}
%%%%%%%%%%%%%%%%%%%%%%%%%%%%%%
%
Equations (\ref{eq:F}), (\ref{eq:JCPmJCPv}), (\ref{eq:Rdef})
and the equalities $F_2 = D^2_2$ and $F_3 = D^2_3$ 
proven above, together with the equalities  
$D_{32}-D_{12} = \Delta m^2_{31}$ and 
$D_{23}-D_{13} = \Delta m^2_{21}$, imply that $R = F_1$.
This completes the proof of the result reported 
in \cite{Krastev:1988yu} and given 
in eqs. (\ref{eq:JCPmJCPv}) and (\ref{eq:F}).

 Two comments are in order.
First, the function 
$F(\theta_{12},\theta_{13},\Delta m^2_{21},\Delta m^2_{31},A)$, 
as determined in eqs. (\ref{eq:F}) is positive. 
Indeed, it follows from eqs. (\ref{eq:D2sqrF2}) and 
(\ref{eq:D3sqrF3}) that the functions $F_2$ and 
$F_3$ are positive. For $A = 0$, 
we have also $F_1D_{12}\, D_{13}\,D_{23}\,D_{32} > 0$.
One can show that this inequality holds also for 
$A \neq 0$, which leads to $F > 0$. 
This implies that the 
rephasing invariants in vacuum and in 
matter, $J_{\rm CP}$ and $J^m$, 
have the same sign:
%%%%%%%%%%%%%%%%%%%%%%%%%%%%%
\begin{equation}
{\rm sgn}\big(J^m\big) = {\rm sgn}\big(J_{\rm CP}\big)\,.
\label{eq:sgnJCP}
\end{equation}
%%%%%%%%%%%%%%%%%%%%%%%%%%%
%
This result is valid both for neutrino mass spetra 
with normal ordering ($\Delta m^2_{31} > 0$) 
and with inverted ordering ($\Delta m^2_{31} < 0$).

 Second, the function  
$F(\theta_{12},\theta_{13},\Delta m^2_{21},\Delta m^2_{31},A)$
in eq. (\ref{eq:JCPmJCPv}) has different equivalent 
representations. This should be clear from the fact that 
%%%%%%%%%%%%%%%%%%%%%%%%%
\begin{align}
\nonumber
J_{\rm CP}(J^m) & =  {\rm Im}\left(
\left(U^{(m)}_{\mu 2}\right)^\ast 
U^{(m)}_{e2}\,U^{(m)}_{\mu 3} \left(U^{(m)}_{e3}\right)^\ast 
\right) \\
\nonumber
& =\,{\rm Im}\left( 
\left(U^{(m)}_{\mu 3}\right)^\ast U^{(m)}_{e3}
\,U^{(m)}_{\mu 1}\left(U^{(m)}_{e1}\right)^\ast
\right) \\
& = {\rm Im}\left( 
U^{(m)}_{\mu 2} \left(U^{(m)}_{e2}\right)^\ast
\,\left(U^{(m)}_{\mu 1}\right)^\ast U^{(m)}_{e1} 
\right) = ...\,,
\label{eq:JCPvm2}
\end{align}
%%%%%%%%%%%%%%%%%%%%%%%%
%
and the derivation presented above. Indeed, 
we can use the second or the third form of $J_{\rm CP}$ ($J^m$) 
in eq. (\ref{eq:JCPvm2}) to obtain the relation 
given in eq. (\ref{eq:JCPmJCPv}).
The function $F$ thus derived will differ in form from, but will be 
equal to, the function $F$ defined in eqs. (\ref{eq:F}) - (\ref{eq:F3}).

 It follows from eqs. (\ref{eq:JCPmJCPv}) 
and (\ref{eq:JCPmJCPv2}) that
%%%%%%%%%%%%%%%%%%%
\be
\dfrac{J^m}{J_{\rm CP}}=
F(\theta_{12},\theta_{13},\Delta m^2_{21},\Delta m^2_{31},A) = 
\tilde{F} = \frac{\Delta m^2_{12}\,\Delta m^2_{23}\,\Delta m^2_{31}}
{\Delta M^2_{12}\,\Delta M^2_{23}\,\Delta M^2_{31}}\,,
\label{eq:FtildeF}
\ee
%%%%%%%%%%%%%%%%%%%%%%%%
% 
i.e., that the function 
$F(\theta_{12},\theta_{13},\Delta m^2_{21},\Delta m^2_{31},A)$ 
found in \cite{Krastev:1988yu} 
is another representation of the function $\tilde{F}$ 
found in \cite{Naumov:1991ju}. The functions $F$ and $\tilde{F}$ 
have very different forms. Nevertheless, as we have verified, 
they coincide numerically. This is illustrated in Fig. \ref{Fig1}
where we show the functions $F$ (eq. (\ref{eq:F})), 
$\tilde{F}$ (eq. (\ref{eq:tildeF})) and the difference 
$(F - \tilde{F})$ versus $A/\Delta m^2_{21}$.
We used the analytic expressions for 
$M^2_i$, $i=1,2,3$, in terms of $m^2_1$, $A$ and the neutrino 
oscillation parameters
$\Delta m^2_{31}$,  $\Delta m^2_{21}$, $\theta_{12}$ and $\theta_{13}$
derived in \cite{Zaglauer:1988gz}. 
It should be clear from eq. (\ref{eq:Hm}) that, 
as we have already discussed,  
in the parametrisation (\ref{eq:PMNSKP1}) employed in 
\cite{Krastev:1988yu} the mass parameters $M^2_i$, $i=1,2,3$, 
do not depend on $\theta_{23}$ and $\delta$.
In Fig. \ref{Fig1}, the neutrino oscillation parameters 
on which the functions $F$ and $\tilde{F}$ depend 
were set to their best fit values 
found in the global analysis of the neutrino 
oscillation data 
% performed 
in \cite{Esteban:2016qun} in the cases 
of NO and IO neutrino mass spectra.

As is suggested by Fig. \ref{Fig1} and we have 
commented earlier, our numerical results 
show that the function $F$ is positive. 
% %%%%%%%%%%%%%%%%%%%%%%%%%%%
% %

 The function $F$ in eq. (\ref{eq:JCPmJCPv}), 
as we have remarked earlier,  
does not depend on $\theta_{23}$ and $\delta$. 
This implies that the ratio 
%%%%%%%%%%%%%%%%%%%%%%%%%%%%%%%%%%%%%%
\begin{equation}
\dfrac{J^m}{\sin2\theta_{23}\sin\delta} =
F\,\dfrac{J_{\rm CP}}{\sin2\theta_{23}\sin\delta}
= \dfrac{1}{8}\,F\,\cos\theta_{13}\,
\sin 2\theta_{12}\,\sin 2\theta_{13}\,,
 \label{eq:JCPmds2th23sd}
\end{equation}
%%%%%%%%%%%%%%%%%%%%%%%%%%%%%%%%%%%
%
does not depend on $\theta_{23}$ and $\delta$.

From  eqs. (\ref{eq:F}),  (\ref{eq:D2sqrF2}), (\ref{eq:D3sqrF3}) 
and (\ref{eq:FUejUmej}),
using  $U^m_{e1} = U_{e1}D_{21}D_{31}/D_1$ 
we find a new expression for the function $F_1$ as well:
%%%%%%%%%%%%%%%%%%%%%%%%%%%%%%%%%%%%%%
\begin{equation}
F_1 = \dfrac{D_2\,D_3}{D_1}\,D_{21}\,D_{31}\,.
\label{eq:F1v2}
\end{equation}
%%%%%%%%%%%%%%%%%%%%%%%%%%%%%%%%%%%
%

%%%%%%%%%%%%%%%%%%%%%%%%%%%
%
\section{The Case of Antineutrino Mixing in Matter} 
% 
%%%%%%%%%%%%%%%%%%%%%%%%%%%%%%%%%%

In the preceding Sections we have focused primarily on the
mixing and oscillations in matter of flavour neutrinos. 
In this Section we will discuss briefly the case 
of mixing and  oscillations in matter of flavour 
antineutrinos. 

 In ordinary matter (of, e.g., the Earth, the Sun)
the mixing of antineutrinos in matter differs from the 
mixing of neutrinos in matter as a consequence of the fact 
that ordinary matter is not charge conjugation invariant: 
it contains protons, neutrons and electrons, but does 
not contain their antiparticles. This causes  
CP and CPT violating effects in the mixing and oscillations 
of neutrinos in matter \cite{Lang87}. As a consequence,  
the neutrino and antineutrino mixing angles, as well the 
masses of the respective neutrino mass-eigenstates,  
in matter differ. The expressions for 
the antineutrino mixing angles in matter, $\overline{\theta}_{ij}$,
the neutrino masses in this case, $\overline{M}_k$, and 
the corresponding $J$-factor, 
$\overline{J}{}^m$, can be obtained from 
those corresponding to neutrino mixing in matter, 
as is well known, by replacing the potential 
$A$ with $(-A)$.

 Since the derivations of the results given 
in eqs. (\ref{eq:FtildeF}) - (\ref{eq:F1v2})
do not depend on the sign of the matter term $A$, these results are valid 
also for mixing of antineutrinos in matter and for 
oscillations of antineutrinos $\overline{\nu}_l$ in matter with 
constant density. Thus, we have:
%%%%%%%%%%%%%%%%%%%
\begin{align}
\label{eq:barJCPmJCPv}
& \overline{J}{}^m = J_{\rm CP}\,
\overline{F}(\theta_{12},\theta_{13},\Delta m^2_{21},\Delta m^2_{31},A)\,,\\
& \overline{F}(\theta_{12},\theta_{13},\Delta m^2_{21},\Delta m^2_{31},A) = 
F(\theta_{12},\theta_{13},\Delta m^2_{21},\Delta m^2_{31},-\,A)\,,
\label{eq:barFF}
\end{align}
%%%%%%%%%%%%%%%%%%%%%%%%
%%%%%%%%%%%%%%%%%%%%%%%%%
\be
\overline{J}{}^m =  {\rm Im}\left( \left(\overline{U}{}^m_{e2}\right) 
\left(\overline{U}{}^m_{\mu 3}\right) \left(\overline{U}{}^m_{e3}\right)^\ast 
\left(\overline{U}{}^m_{\mu 2}\right)^\ast \right) = 
\frac{1}{8}\,\cos\overline{\theta}{}^m_{13}
\sin2\overline{\theta}{}^m_{12}\,\sin 2\overline{\theta}{}^m_{23}\,
\sin2\overline{\theta}{}^m_{13}\,\sin\overline{\delta}{}^m\,,
\label{eq:barJCPvm}
\ee
%%%%%%%%%%%%%%%%%%%%%%%%%%%%%%
%
where 
$\overline{U}{}^m_{lj}$ are the elements 
of the antineutrino mixing matrix in matter 
$\overline{U}{}^m$, $\overline{\delta}{}^m$ is the 
Dirac phase present in $\overline{U}{}^m$, 
and $\overline{\theta}{}^m_{ij}$ are the 
antineutrino mixing angles in matter.
We also have:
%%%%%%%%%%%%%%%%%%%%%%%%%%%%%%%%%
\begin{equation}
\sin2\overline{\theta}{}^m_{23}\,\sin\overline{\delta}{}^m
= \sin2\theta_{23}\,\sin\delta\,,
\label{eq:s2thbar23mdm}
\end{equation}
%%%%%%%%%%%%%%%%%%%%%%%%%%%%%%%
%%%%%%%%%%%%%%%%%%%%%%%%%%%%%%%%%
\begin{equation}
\overline{F} = \dfrac{\overline{U}{}^m_{e1}\,\overline{U}{}^m_{e2}\,
\overline{U}{}^m_{e3}}{U_{e1}\,U_{e2}\,U_{e3}} 
= \frac{\Delta m^2_{12}\,\Delta m^2_{23}\,\Delta m^2_{31}}
{\Delta \overline{M}{}^2_{12}\,\Delta \overline{M}{}^2_{23}\,
\Delta \overline{M}{}^2_{31}}\,,
\label{eq:FUejbarUmej}
\end{equation}
%%%%%%%%%%%%%%%%%%%%%%%%%%%%%%
%
 with $\overline{M}{}^2_{ij} = \overline{M}{}^2_i - \overline{M}{}^2_j$.
From the exact relations (\ref{eq:s2th23mdm}) and (\ref{eq:s2thbar23mdm}) 
we get  
%%%%%%%%%%%%%%%%%%%%%%%%%%%%%%%%%
\begin{equation}
\sin2\theta^m_{23}\,\sin\delta^m
= \sin2\overline{\theta}{}^m_{23}\,\sin\overline{\delta}{}^m 
= \sin2\theta_{23}\,\sin\delta\,,
\label{eq:s2th23mdmbar}
\end{equation}
%%%%%%%%%%%%%%%%%%%%%%%%%%%%%%%
%
while eqs. (\ref{eq:FtildeF})
and (\ref{eq:FUejbarUmej}) imply:
%%%%%%%%%%%%%%%%%%%%%%%%%%%%%%%%
\begin{equation}
\overline{F} = F\,
\frac{\Delta M^2_{12}\,\Delta M^2_{23}\,\Delta M^2_{31}}
{\Delta \overline{M}{}^2_{12}\,\Delta \overline{M}{}^2_{23}\,
\Delta \overline{M}{}^2_{31}}\,,
\label{eq:barFF2}
\end{equation}
%%%%%%%%%%%%%%%%%%%%%%%%%%%%%%%%%%
%
Finally, as in the neutrino mixing in matter case, 
the equalities 
%%%%%%%%%%%%%%%%%%%%%%%%
\begin{align} 
\label{eq:thm23eqth23anu}
& \overline{\theta}{}^m_{23} = \theta_{23}\,,~~~~0 < \theta_{23},
\overline{\theta}{}^m_{23}\leq \pi/2\,,\\
\label{eq:dmdanu}
&\overline{\delta}{}^m = \delta\,,
\end{align}
%%%%%%%%%%%%%%%%%%%%%%%%%%%
%
although not exact, represent an excellent 
approximations for the ranges of values of 
 $A/\Delta m^2_{21}$ relevent for the 
T2K (T2HK), T2HKK, NO$\nu$A and DUNE
neutrino oscillation experiments.

\section{Summary} 
%
%%%%%%%%%%%%%%%%%%%%%%%%%%%%%
 
 In the present article 
we have analysed aspects of 3-neutrino mixing 
in matter and of CP and T violation in 3-flavour neutrino  
oscillations in vacuum and 
in matter with constant density. 
The analyses have been performed in the parametrisation 
of the PMNS neutrino mixing matrix $U_{\rm PMNS}\equiv U$
specified in eq. (\ref{eq:PMNSKP1}) and 
introduced in \cite{Krastev:1988yu}.
However, as we have shown, the results obtained 
in our study are valid (in some cases with trivial modifications) 
also in the standard parametrisation of the PMNS matrix 
(see, e.g., \cite{PDG2017}).
 
 Investigating the case of  3-neutrino mixing 
in matter with constant density we have 
derived first the relations 
$\theta^m_{23} = \theta_{23}$ and $\delta^m = \delta$,
$\theta_{23}$ ($\theta^m_{23}$) and $\delta$ ($\delta^m$)
being respectively the atmospheric neutrino 
mixing angle and the Dirac CP violation phase  
in vacuum (in matter) present in the PMNS 
neutrino mixing matrix. 
Performing a detailed 
numerical analysis we have shown that 
although these equalities are not exact, 
they represent excellent approximations
for the ranges of values of $A/\Delta m^2_{21} < 30$ relevent for the 
T2K (T2HK), T2HKK, NO$\nu$A and DUNE 
neutrino oscillation experiments,
the deviations from each of the two 
relations not exceeding respectively $1.3\times 10^{-2}$ and 
$1.3\times 10^{-3}$ (Figs. \ref{Fig3} and \ref{Fig32})).
Similar conclusion is  valid for the corresponding parameters 
 $\overline{\theta}{}^m_{23}$ and  $\overline{\delta}{}^m$
in the case of mixing of antineutrinos 
(Figs. \ref{Fig33} and \ref{Fig34})).

We have derived next the relation  
$\sin2\theta^m_{23} \sin\delta^m = \sin2\theta_{23} \sin\delta$, 
and have shown numerically that it is exact 
(Fig. \ref{Fig2}). The relation is well known 
in the literature (see \cite{Toshev:1991ku,Freund:2001pn}).
We have presented a new derivation of this result.
Using the indicated relation 
 and the fact that the deviations of
$\theta^m_{23}$ from $\theta_{23}$, 
$\epsilon_{23}(A/\Delta m^2_{21})$,
and of $\delta^m$ from 
$\delta$, $\epsilon_{\delta}(A/\Delta m^2_{21})$, 
are small, $|\epsilon_{23}|,|\epsilon_{\delta}| \ll 1$, 
we have derived a relation between 
$\epsilon_{23}$ and $\epsilon_{\delta}$
working in leading order in these two parameters
(eq. (\ref{eq:corrth23d})). It follows from this relation, 
in particular, that for $\theta_{23} = \pi/4$ 
and $\delta \neq 3\pi/2,\pi/2$, 
the leading order matter correction to $\delta$ vanishes, 
while for $\delta = 3\pi/2$ ($\pi/2$) and 
$\theta_{23}\neq \pi/4$, the leading order matter correction 
to $\theta_{23}$ vanishes.
 
  We have discussed further
the relation between the rephasing invariants, 
associated with the Dirac phase 
in the neutrino mixing matrix, which determine 
the magnitude of CP and T violating effects 
in 3-flavour neutrino  oscillations 
in vacuum, $J_{\rm CP}$, and of the T violating effects
in matter with constant density, $J^{m}_{\rm T} \equiv J^{m}$,
obtained in \cite{Krastev:1988yu}:  
$J^{m} = J_{\rm CP}\,F$.
$F$ is a function 
whose explicit form in terms of the squared masses 
in vacuum and in matter of the mass-eigenstate neutrinos, 
of the solar and reactor neutrino mixing angles and 
of the neutrino matter potential (eq. (\ref{eq:F})) was given 
in \cite{Krastev:1988yu}. 
The quoted relation between $J^{m}$ and 
$J_{\rm CP}$ was reported in \cite{Krastev:1988yu}
without a proof. We have presented a derivation of this relation. 
We have shown also that the function 
$F= F(\theta_{12},\theta_{13},\Delta m^2_{21},\Delta m^2_{31},A)$ 
i) is positive, $F > 0$, which implies that 
 $J^{m}$ and $J_{\rm CP}$ have the same sign,
${\rm sgn}(J^m) = {\rm sgn}(J_{\rm CP})$,
and that 
ii) it can have different forms.
We have proven also that the function $F$ as given 
in \cite{Krastev:1988yu} is another representation 
of the so-called called ``Naumov factor'' (Fig. \ref{Fig1}):
$F = \Delta m^2_{12}\Delta m^2_{23}\Delta m^2_{31}
(\Delta M^2_{12}\Delta M^2_{23}\Delta M^2_{31})^{-1}$,
where $\Delta m^2_{ij} = m^2_i - m^2_j$, 
$\Delta M^2_{ij} =  M^2_i - M^2_j$,
$m_i$ and $M_i$, $i=1,2,3$, being the masses 
of the three mass-eigenstate neutrinos in vacuum 
and in matter.

Finally, we have considered briefly 
the case of antineutrino mixing in matter and
have shown that results similar to those 
derived for the mixing of neutrinos in matter are 
valid also in this case.

 The results of the present study contribute to the 
understanding of the neutrino mixing in matter 
and flavour neutrino oscillations 
in matter with constant density, widely 
explored in the literature on the subject.
They could be useful for the studies of neutrino oscillations  
in long baseline neutrino oscillation experiments 
T2K (T2HK), T2HKK, NO$\nu$A and DUNE.

\vspace{0.6cm}
{\bf Acknowledgements.}
This work was supported in part 
by the INFN program on Theoretical Astroparticle Physics (TASP),
by the European Union Horizon 2020 research and innovation programme
under the  Marie Sklodowska-Curie grants 674896 and 690575, 
by the  World Premier International Research Center 
Initiative (WPI Initiative, MEXT), Japan (S.T.P.), 
as well as by the European Research Council under 
ERC grant ``NuMass'' FP7-IDEAS-ERC ERC-CG 617143
(Y.L.Z.).

\end{document}